\documentclass[journal]{IEEEtran}
\usepackage[noadjust]{cite}

\ifCLASSINFOpdf
  
\else
\usepackage[dvips]{graphicx,color}
  
\fi
\usepackage[cmex10]{amsmath}

\usepackage{multirow}
\usepackage{amsmath,amssymb,mathrsfs}
\usepackage{type1cm}

\newtheorem{teiri}{Theorem}
\newtheorem{kei}{Corollary}
\newtheorem{meidai}{Proposition}
\newtheorem{hodai}{Lemma}
\newtheorem{teigi}{Definition}
\newtheorem{rei}{Example}
\newtheorem{chui}{Remark}

\newcommand{\bal}{\begin{align}}
\newcommand{\non}{\nonumber}

\renewcommand{\IEEEQED}{\hfill\IEEEQEDopen}
\newcommand{\Exp}{{\rm E}}

\newcommand{\spr}[1]{{\bf #1}}

\newcommand{\vep}{\varepsilon}
\newcommand{\vph}{\varphi}

\newcommand{\barY}{\overline{Y}}

\renewcommand{\thesection}{\Roman{section}}

\newcommand{\cB}{{\cal B}}
\newcommand{\cC}{{\cal C}}

\newcommand{\cK}{{\cal K}}

\newcommand{\cI}{{\cal I}}

\newcommand{\cL}{{\cal L}} 
\newcommand{\cM}{{\cal M}}
\newcommand{\cP}{{\cal P}}
\newcommand{\cR}{{\cal R}} 
\newcommand{\cS}{{\cal S}}
\newcommand{\cU}{{\cal U}} 
\newcommand{\cV}{{\cal V}}
\newcommand{\cX}{{\cal X}}
\newcommand{\cY}{{\cal Y}}
\newcommand{\cT}{{\cal T}}
\newcommand{\cZ}{{\cal Z}}

\newcommand{\sS}{\spr{S}}

\newcommand{\sU}{\spr{U}}
\newcommand{\sV}{\spr{V}}
\newcommand{\sX}{\spr{X}}
\newcommand{\sY}{\spr{Y}}
\newcommand{\sZ}{\spr{Z}}

\newcommand{\ssu}{\spr{u}}
\newcommand{\ssv}{\spr{v}} 
\newcommand{\ssx}{\spr{x}}
\newcommand{\ssz}{\spr{z}}
\newcommand{\ssy}{\spr{y}}
\newcommand{\sss}{\spr{s}}

\newcommand{\tU}{\tilde{U}} 
\newcommand{\tV}{\tilde{V}} 

\newcommand{\bteiri}{\begin{teiri}}
\newcommand{\eteiri}{\end{teiri}}
\newcommand{\bkei}{\begin{kei}}
\newcommand{\ekei}{\end{kei}}
\newcommand{\brei}{\begin{rei}}
\newcommand{\erei}{\end{rei}}
\newcommand{\bhodai}{\begin{hodai}}
\newcommand{\ehodai}{\end{hodai}}
\newcommand{\bmeidai}{\begin{meidai}}
\newcommand{\emeidai}{\end{meidai}}
\newcommand{\bteigi}{\begin{teigi}}
\newcommand{\eteigi}{\end{teigi}}
\newcommand{\bchui}{\begin{chui}}
\newcommand{\echui}{\end{chui}}
\newcommand{\beq}{\begin{equation}}
\newcommand{\eeq}{\end{equation}}
\newcommand{\beqn}{\begin{eqnarray}}
\newcommand{\eeqn}{\end{eqnarray}}
\newcommand{\beqns}{\begin{eqnarray*}}
\newcommand{\eeqns}{\end{eqnarray*}}

\newcommand{\map}{\vph_n: \cX^n \to \cY^n}
\newcommand{\mapMtoY}{\vph_n: \cM_{M_n} \to \cY^n}
\newcommand{\mapXtoM}{\vph_n: \cX^n \to \cM_{M_n}}

\begin{document}
%
\title{
Wiretap Channels with Causal and
\\Non-Causal
 State Information: Revisited
 %
 %
 %
 %
}
\author{
Te~Sun~Han,~\IEEEmembership{Life Fellow,~IEEE}, Masahide Sasaki
\thanks{T. S. Han and M. Sasaki are with the Quantum ICT Collaboration Center, 
National Institute of Information and Communications Technology (NICT), 
Nukuikitamachi 4-2-1, Koganei, Tokyo 184-8795, Japan 
(email: han@is.uec.ac.jp, psasaki@nict.go.jp).
Copyright (c) 2021 IEEE. Personal use of this material is permitted.
However, permission to use this material for any other purposes must be
obtained from the IEEE by sending a request to pubs-permissions@ieee.org}}
%

%
\maketitle
\begin{abstract} 
 The coding problem for  wiretap channels (WTCs) with  {\em causal} and/or  {\em non-causal} 
 channel state information (CSI) available at the encoder 
 (Alice) and/or the decoder (Bob)  is studied, particularly focusing on
 achievable  secret-message  secret-key (SM-SK) rate pairs
 under the {\em semantic security} criterion. 
One of our main results  
is summarized as Theorem \ref{teiri:2}  on causal inner bounds for  SM-SK rate pairs,
which follows immediately by leveraging the  unified seminal theorem 
for WTCs with {\em non-causal} CSI at Alice that has been recently established by Bunin {\em et al.} \cite{bunin-2019}.
The only thing to do here is just to re-interpret the latter non-causal scheme in a causal manner
by restricting the range of auxiliary random variables appearing  in  non-causal encoding  to a subclass 
of auxiliary random variables for the causal encoder. This technique is referred to as ``plugging."
%
%
Then, we are able to dispense with the block-Markov encoding scheme used in the previous works by Chia and El Gamal \cite{chia-elgamal}, 
Fujita \cite{fujita}, and Han and Sasaki \cite{han-sasaki-c} and then  extend  all the known  results on achievable rates.
%
The other main results include the exact SM-SK capacity region for WTCs with non-causal  CSI at “both” Alice and Bob (Theorem \ref{teiri:3}),
 a ``tighter" causal SM-SK outer bound for state-reproducing coding schemes with CSI at Alice (Proposition  \ref{teiri:yatta1}), 
 and the exact SM-SK capacity region for degraded WTCs with causal/non-causal CSI at  both Alice and Bob (Theorem \ref{teiri:4}).  
%

%
\bigskip
\bigskip
\end{abstract}



%

\section{Introduction}\label{introduction1}

In this paper  we address the coding problem for a wiretap channel (WTC) with {\em causal/non-causal} channel state information (CSI) available at the encoder (Alice) and/or the decoder (Bob).
 The intriguing concept of WTC and secret message  (SM) transmission through the WTC 
 originates in Wyner \cite{wyner-wire} (without CSI)
 under the {\em weak} secrecy criterion. This was  then  extended to a wider class of 
 WTCs by Csisz\'ar and K\"orner \cite{csis-kor-3rd} to provide  the more tractable framework. Indeed, 
these landmark papers have offered the fundamental basis for  a diversity  of 
subsequent extensive researches.

Early works  include 
Mitrpant, Vinck and Luo \cite{mitrpant},  Chen and Vinck \cite{chen-vinck},
and  Liu and Chen \cite{liu-chen}  that have studied  
the  capacity-equivocation tradeoff  for degraded WTCs with {\em non-causal} CSI to 
establish inner and/or outer bounds on the achievable region.
  Subsequent developments in this direction 
with {\em non-causal} CSI can be found also in 
%
 Boche and Schaefer \cite{boche}, Dai and Luo \cite{dai-luo}, 
 etc., which are mainly concerned with
 the problem of SM transmission over the WTC.

On the other hand,
   Khisti, Diggavi and Wornell \cite{khisti-wornell}  and Zibaeenejad \cite{zibae}  addressed the  problem of 
   secret key (SK) agreement over the WTC
    with {\em non-causal} CSI
  at Alice (and Bob), and tried to give the {\em exact} key-capacity formula.

%
Prabhakaran {\em et al.}
\cite{prabhakaran}
%
 studied an achievable tradeoff between 
SM and SK rates over the WTC with non-causal CSI, 
deriving a benchmark inner bound on the SM-SK capacity region under the weak secrecy criterion.
Subsequently, heavily based on the work of 
Goldfeld  {\em et al.} \cite{goldfeld}, Bunin  {\em et al.}  \cite{bunin, bunin-2019}
have  improved  \cite{prabhakaran} by explicitly leveraging  the superposition coding to obtain 
%
a unifiying formula (cf. Theorem \ref{teiri:1})
 for  inner bounds 
on the SM-SK capacity region under the semantic secrecy (SS) criterion for WTCs with non-causal CSI at Alice, from which ``all"
  the  typical previous results can be derived.
Thus,  \cite{bunin, bunin-2019} are regarded currently as establishing the best known achievable rate pairs
 with {\em non-causal} CSI at Alice.

%
%
 
 The key idea in \cite{bunin, bunin-2019} (which are substantially due to \cite{goldfeld})  is to invoke 
the {\em likelihood encoder} (cf. Song {\em et al.}
 \cite{eva}) together with the   
  {\em soft-covering lemma}  (cf. Cuff \cite{cuff-s}) 
  \footnote{This is the notion to denote  the achievability part of  {\em resolvability} \cite{ver-han}. }
on the basis of 
  two layered superposition coding scheme (cf. 
\cite{prabhakaran}, \cite{goldfeld}), which makes it  possible to guarantee  the 
  {\em semantically secure} (SS) information transmission.
  This is one of the strongest ones among  various security criteria.

  In contrast to extensive studies on WTCs with ``non-causal" CSI mentioned above, 
  there have been less number of literatures on WTCs with ``causal" CSI.
  %
To our best knowledge, we can list typically  a few causal papers including  
  Chia and El Gamal \cite{chia-elgamal}, Fujita \cite{fujita}, and Han and Sasaki \cite{han-sasaki-c}.
They  are concerned only with SM rates but not with SK rates.

  A prominent feature common in these papers is to leverage the block-Markov encoding to invoke the Shannon cipher 
   \cite{shannon-secrecy} (Vernam's one-time pad cipher). Although there still remain many open problems, possible extensions/generalizations in this direction do not seem to be very fruitful or may be even formidable.

 Fortunately, however, to solve these problems we can fully exploit, as they are,  all the {\em non-causal} techniques/concepts
 as developed in Bunin {\em et al.} \cite{bunin-2019} to derive the {\em causal} version of it.
  The only  thing to do here is simply to restrict the range of auxiliary random variables $(U, V)$'s intervening in 
  \cite[Theorem 1]{bunin-2019} (said to be {\em non-causally} achievable) to a subclass  of auxiliary random variables 
 $(U, V)$'s (said to be {\em causally} achievable). 
 Then, it suffices to notice only  that
 the encoding scheme given in  \cite{bunin-2019} can be carried out, as it is,  in a causal way.
 This process may be termed   ``plugging"  of causal WTCs into non-causal WTCs.

  Thus, it is not necessary to give a separate proof to establish the causal version (Theorem \ref{teiri:2}) 
  in this paper.
  The merits of this approach
   for proof are to inherit all the advantages in  \cite{bunin-2019} to our causal version.
For example, 
 the first one is to  inherit the SS property  as established in \cite{bunin-2019};
 the second one is to enable us, without any extra arguments,  to interpret regions of 
SM-SK achievable rate pairs in  \cite{bunin-2019} as those valid also in Theorem \ref{teiri:2};
  the third one is 
  to enable us to dispense with the 
 involved block-Markov 
 encoding scheme   (cf. \cite{chia-elgamal}, \cite{han-sasaki-c});
  the fourth one is 
  that all the results as established in \cite{chia-elgamal}, \cite{fujita},  \cite{han-sasaki-c}
 follow immediately from  Theorem \ref{teiri:2};  
 the fifth one is to be able to derive, in  a straightforward manner, a variety of  novel causal inner bounds  
 on the SM-SK capacity region; 
 %
 the sixth one is, as a by-product,  to enable us to exactly
 determine the general formula for the SM-SK  capacity region for WTCs with {\non non-causal} CSI 
 available at both Alice and Bob (Theorem \ref{teiri:3}).
 
 Furthermore,  the arguments that have been  used to derive 
 Theorems
  \ref{teiri:3} and \ref{teiri:2}  can be further exploited to solve harder problems such as deriving 
 a ``tighter" causal outer bound (Theorem \ref{teiri:yatta1}) and finding the causal/non-causal SM-SK capacity region for degraded WTCs
 (Theorem \ref{teiri:4}).

 The present paper is organized as follows.

In Section \ref{sec-non-causal-WTC}, we give the problem statement  
 as well as the necessary notions and notation,
all of which are borrowed from \cite{bunin-2019} along with Theorem \ref{teiri:1} 
with non-causal CSI at Alice. They are used in 
the next sections.

In particular, in Section \ref{sec:non-alice-bob}, we give
Theorem \ref{teiri:3} to demonstrate
the general formula for the exact ``non-causal" SM-SK capacity region
when the state information is available at both  Alice and Bob.

In Section \ref{intro-ge-cs1}, 
we give the   proof of Theorem \ref{teiri:2} 
for WTCs with causal CSI at Alice by using the argument of ``plugging," which is to put the causal scenario into
the non-causal scenario, thereby enabling us to produce  a diversity of causal inner bounds
 in Section \ref{sec-application-causal-WTC}.

%
In Section \ref{sec-application-causal-WTC}, 
we develop Theorem \ref{teiri:2} 
for each of Case 1) $\sim$ Case 4) to obtain a new class of 
inner bounds of SM-SK  achievable rate pairs
for WTCs with causal CSI at Alice. Here, it is also shown that all the results as established 
in  \cite{chia-elgamal},  \cite{fujita}, and  \cite{han-sasaki-c} can be derived as special cases of 
Theorem \ref{teiri:2}. Furthermore, in this section we give 
%
Proposition \ref{teiri:yatta1}
for state-reproducing coding schemes (with causal CSI at Alice) to derive an  SM-SK outer bound, which is 
paired with Proposition \ref{prop:1} (inner bound).

In Section \ref{sec-sececy-region}, 
we establish the exact SM-SK  capacity region with {\em causal/non-causal} CSI available at  both Alice  and Bob (Theorem \ref{teiri:4} for degraded WTCs),
which is the first solid result from the viewpoint of ``causal" SM-SK capacity regions.
%

In Section \ref{sec-revisited-remark}, we conclude the paper with several remarks.

Finally, in Appendix \ref{addA}, we give an elementary proof of the soft-covering lemma
that plays the key role in \cite{bunin-2019, goldfeld}. 
Also, the proof of Remark \ref{chui:sune3}
on typical causal  inner bounds is given in Appendix \ref{addB}.

\section{Wiretap Channel with  Non-Causal CSI}\label{sec-non-causal-WTC}

\begin{figure}[htbp]
\begin{center}
\includegraphics[width=80mm]{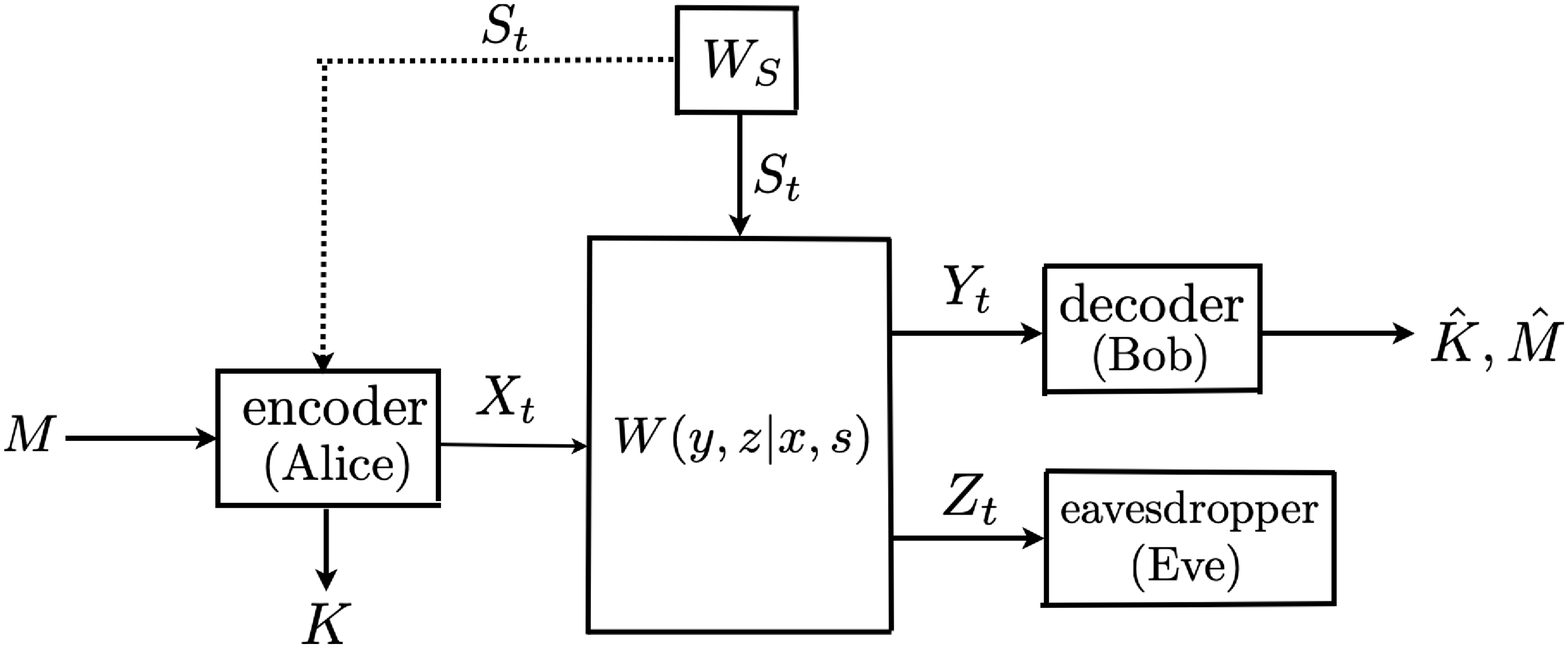}
\end{center}
\caption{WTC with CSI available only at Alice ($t=1,2,\cdots, n$).
          }
\label{fig1}
\end{figure}

In this section,  we recapitulate the seminal work for wiretap channels with ``{\em non-causal}" channel
state information (CSI) available at the encoder (Alice)  as in  Fig. \ref{fig1},  which was recently 
established  by the group of Bunin, Goldfeld, Permuter, Shamai, Cuff and Piantanida \cite{bunin-2019}.
For the reader's convenience, we   repeat here their   notions  and key result as they are.
Leveraging them, we derive the ``{\em causal}" counterparts in Section \ref{intro-ge-cs1}. 

{\rm  II. A}: {\em Problem Statement}

\smallskip

Let $\cS, \cX, \cY, \cZ$ be finite sets and $\cS^n, \cX^n, \cY^n, \cZ^n$
be the $n$ times product sets. We let $(\cS, \cX, \cY, \cZ, W_S,$ $ W_{YZ|SX})$ denote a
discrete stationary and  memoryless WTC with ``non-causal" stationary memoryless CSI $S$ 
available at the encoder, 
where $W_{YZ|SX}: \cS\times \cX\to \cP(\cY\times \cZ)$
\footnote{$\cP(\cal D)$ denotes the set of all probability distributions on the set ${\cal D}$.
Also, we use $p_U$ to denote the probability distribution of a random variable $U$.
Similarly, we use $p_{U|V}$ to denote the conditional probability distribution for $U$ given $V$.
}
 is 
the transmission probability distribution (under  state $S$)
with input $X$ at Alice, and  outputs $Y$ at Bob and  $Z$ at Eve, while
 $W_S$ is the probability distribution of state variable $S$.
A  state sequence $\sss\in \cS^n$ is sampled in an i.i.d. manner according to $W_S$
and revealed in  a non-causal fashion to Alice. Independently of the observation of $\sss$, 
Alice chooses a message $m$ from the set
\footnote{For  integers $r\le l$, $[r:l]$ denotes $\{r, r+1, \cdots, l-1, l\}$.} 
$[1:2^{nR_M}]\ (R_M\ge 0)$ and maps  the pair $(\sss, m)$ into a channel input sequence $\ssx\in \cX^n$ and a key index 
$k\in [1:2^{nR_K}]\  (R_K\ge 0$; the mapping may be stochastic).
The sequence $\ssx$ is transmitted over the WTC under state $\sss$.
The output sequences $\ssy\in \cY^n$ and $\ssz\in \cZ^n$ are observed by the legitimate receiver (Bob) and the eavesdropper (Eve),
respectively. Based on $\ssy$, Bob  produces the pair $(\hat{k}, \hat{m})$ as an estimate of $(k,m)$.
Eve  maliciously attempts to decipher the SM-SK rate pair from $\ssz$ as much as possible.
The random variables corresponding to $\sss, \ssx, \ssy, \ssz, m, k$ may be  denoted by
 $S^n, X^n, Y^n, Z^n$ (or also $\sS, \sX, \sY, \sZ )$, $M, K$;  respectively.

The following Definitions \ref{teigi:1} $ \sim$ \ref{teigi:6} are borrowed from \cite{bunin-2019}.
\bteigi[Non-causal code]\label{teigi:1}
{\rm
An $(n, R_M, R_K)$-code $c_n$ for the WTC with ``non-causal"  CSI  at Alice and  message set 
$\cM_n\stackrel{\Delta}{=}[1:2^{nR_M}]$ and  key set $\cK_n\stackrel{\Delta}{=}[1:2^{nR_K}]$ is
a pair  of functions $(f_n, \phi_n)$ such that

  
 1) $f_n: \cM_n\times \cS^n \to \cP( \cX^n\times \cK_n)$,
 
 
 2) $\phi_n: \cY^n\to \cM_n\times\cK_n$,
 
 where $f_n$  is a  stochastic function.
 }
\eteigi

The performance of the code $c_n$ is evaluated in terms of  its rate pair $(R_M, R_K)$, 
the maximum decoding error probability, the key uniformity and independence metric, 
and SS metric as follows:
\bteigi[Error Probability]\label{teigi:2}
{\rm
The error probability of an $(n, R_M, R_K)$-code $c_n$ is
\beq\label{eq:1}
e(c_n) \stackrel{\Delta}{=} \max_{m\in \cM_n} e_m(c_n),
\eeq
where, for every $m\in \cM_n$,
\beq\label{eq:2}
e_m(c_n) \stackrel{\Delta}{=} \Pr\{(\hat{M}, \hat{K}) \neq (m, K) | M=m\}
\eeq
with the decoder output $(\hat{M}, \hat{K})\stackrel{\Delta}{=}\phi_n(Y^n)$.
}
\eteigi
\bteigi[Key Uniformity and Independence Metric]\label{teigi:3}
{\rm
The key uniformity and independence (from the message) metric under 
$(n, R_M, R_K)$-code $c_n$ is 
\beq\label{eq:3}
\delta (c_n) \stackrel{\Delta}{=}  \max_{m\in \cM_n}\delta_m(c_n),
\eeq
where, for every $m\in \cM_n$,
\beq\label{eq:4}
\delta_m(c_n) \stackrel{\Delta}{=}  ||p^{(c_n)}_{K|M=m} -p_{\cK_n}^{(U)}||_{{\sf TV}},
\eeq
and $p^{(c_n)}$ denotes the joint probability distribution  over the WTC induced by the code $c_n$; 
$p_{\cK_n}^{(U)}$ is the uniform distribution over $\cK_n$, and $||\cdot||_{{\sf TV}}$
denotes the total variation.
}
\eteigi
\bteigi[Information Leakage and SS-Metric]\label{teigi:4}
{\rm
The information leakage to Eve under $(n, R_M, R_K)$-code $c_n$ and message 
distribution $p_M \in \cP(\cM_n)$ is $\ell(p_M, c_n)
\stackrel{\Delta}{=}  I_{p^{(c_n)}}(M, K;\sZ),$
where $I_{p^{(c_n)}}$ denotes the mutual information with respect to the joint probability $p^{(c_n)}$.
The SS-metric with respect to $c_n$ is 
\beq\label{eq:5}
\ell_{{\sf Sem}}(c_n) \stackrel{\Delta}{=} \max_{p_M \in \cP(\cM_n)}\ell (p_M, c_n).
\eeq
}
\eteigi
\bteigi[Achievability]\label{teigi:5}
{\rm
A pair $(R_M, R_K)$ is called an   SM-SK achievable rate pair for the WTC with non-causal  CSI at Alice, if
for every $\epsilon >0$ and sufficiently large $n$ there exists an $(n, R_M, R_K)$-code $c_n$ with
\beq\label{eq:6}
\max [e(c_n), \delta (c_n),  \ell_{{\sf Sem}}(c_n)] \le \epsilon.
\eeq
}
\eteigi
\bteigi[Non-causal SM-SK capacity region]\label{teigi:6}
{\rm
Throughout in this paper we use the following notation.
The SM-SK capacity region  of  the WTC with non-causal  CSI at Alice,
denoted by $\cC_{\mbox{{\scriptsize\rm NCSI-E}}}$
\footnote{E denotes Encoder=E and  N of NCSI denotes Non-causal=N.},  is
the set of all  SM-SK achievable rate pairs.  Furthermore, 
the supremum of the projection of $\cC_{\mbox{{\scriptsize\rm NCSI-E}}}$ on the $R_M$-axis, denoted by
 $C^{{\scriptsize\rm M}}_{\mbox{{\scriptsize\rm NCSI-E}}}$,
 is called the SM capacity,
whereas 
the supremum of the projection of $\cC_{\mbox{{\scriptsize\rm NCSI-E}}}$ on the $R_K$-axis is called the SK capacity,
denoted by
 $C^{{\scriptsize\rm K}}_{\mbox{{\scriptsize\rm NCSI-E}}}$.

} 
\eteigi

\medskip

{\rm II. B}: {\em Wiretap Channel with Non-causal  CSI at Alice}

\smallskip

We can now
 describe the unifying key theorem of 
  \cite{bunin-2019}.
Let $\cU, \cV$ be finite sets and let $U, V$ be  random variables 
taking values in $\cU, \cV$, respectively, where $U, V, S, X$ may be correlated.
Define  joint  probability
 distributions $p_{YZXSUV}$ on $\cY\times \cZ \times \cX  \times \cS \times \cU \times \cV$
 (said to be {\em non-causally achievable})
 so that $UV\to SX\to YZ$ forms a Markov chain
 \footnote{We may use $UV, SX, UV$ instead of $(U,V), (S,X), (U, V)$, and so on, for notational simplicity.}
 and 
\beq\label{eq:non1}
p_S=W_S, \ \ p_{YZ|SX}=W_{YZ|SX}.
\eeq
Notice here  that,
in view of (\ref{eq:non1}), such a distribution $p_{YZXSUV}$ is  specified by giving 
the marginal $p_{SUV}$ (input),
so we may use $p_{SUV}$ in short  instead of $p_{YZXSUV}$.
Define $\cR_{\sf in}(p_{SUV})$ to be the set of all nonnegative rate pairs $(R_M, R_K)$ satisfying
the rate constraints:
\beqn
R_M &\le& I(UV; Y)-I(UV; S),\label{eq:non2}\\
R_M+R_K &\le& I(V;Y|U)-I(V;Z|U) \non\\
& &\qquad - [I(U;S)-I(U;Y)]^+,
\label{eq:non3}
\eeqn
where $[x]^+ = \max (x, 0)$ and $I(\cdot;\cdot), I(\cdot;\cdot|\cdot)$ denotes the (conditional) mutual information.

With these definitions, Bunin {\em et al.}
  \cite{bunin-2019} 
 have established 
the following non-causal inner bound:
%

\bteiri[Non-causal SM-SK  inner bound]\label{teiri:1}
\beq\label{eq:non-4}
\cC_{\mbox{{\scriptsize\rm NCSI-E}}} \supset \cR_{\sf in}^{\sf N}
 \stackrel{\Delta}{=}
  \bigcup_ {{\sf N:}p_{SUV}} \cR_{\sf in}(p_{SUV}),
\eeq
where  the union  is taken over all ``non-causally" achievable probability 
distributions
 $p_{SUV}$'s.
 Here,
the cardinalities of $U, V$ may be restricted 
to
$|\cU|\le (|\cX|-1) |\cS|+3$ and $ |\cV|\le (|\cX|-1)^2|\cS|^2+3(|\cX|-1)|\cS|+2$.
\eteiri

%
%
%
%

%
%
%
\bchui\label{chui:naotta}
{\rm
In particular, in Section \ref{sec:non-alice-bob}, the inner bound given by Theorem \ref{teiri:1} is shown to be optimal 
when the state information is available  at both Alice and Bob.
\IEEEQED
}
\echui
\bchui\label{chui:opt-2}
It should be emphasized also that 
the technical crux of the paper 
 \cite{bunin-2019}
 (due to \cite{goldfeld})
  is based on the soft covering lemma
\footnote{A ``stronger" version of the soft covering lemma is given in \cite{cuff-s}, although it is actually 
not necessary to  prove Theorem \ref{teiri:1}.}, 
which is summarized  as
\bhodai[{\rm \cite[Lemma 4]{goldfeld}}]\label{hodai:naotta}
{
Let $W:\cU\times \cV\to \cS$ be the memoryless channel induced
 by  joint probability distribution $p_{SUV}$, and set, with $L_n=2^{nR_1}$ and  $N_n=2^{nR_2}$,
\beq\label{eq:mon1}
q_S^n(\sss) = \frac{1}{L_nN_n}\sum_{i=1}^{L_n}\sum_{j=1}^{N_n}W(\sss|\ssu_i, \ssv_{ij}). 
\eeq
Then, 
for any small $\vep>0$ and for all sufficiently large $n$, it holds that
\beq\label{eq:monk1}
\Exp D(q^n_S||p^n_S) \le \vep,
\eeq
provided that rate constraints $R_1> I(U;S), R_1+R_2> I(UV; S)$ are satisfied,
where $D(Q||P)$ denotes the Kullback-Leibler divergence between $Q$ and $P$, 
and $p^n_S(\sss)$  indicates the probability of i.i.d. $\sss=(s_1, s_2, \cdots, s_n)$
and $\Exp$ denotes the  expectation over all random codewords $\ssu_i, \ssv_{ij}$ of Codebook $\cB_n$ 
 as given later in Section \ref{intro-ge-cs1}.
%
\IEEEQED
}
\ehodai

Although  in this paper  we do not  use  explicitly this lemma,   
in view of its importance,  it would be worthy  of giving  a separate elementary proof,
which is stated in Appendix \ref{addA}.
\echui


\section{Capacity region with non-causal CSI at Alice and Bob}\label{sec:non-alice-bob}

\begin{figure}[htbp]
\begin{center}
\includegraphics[width=80mm]{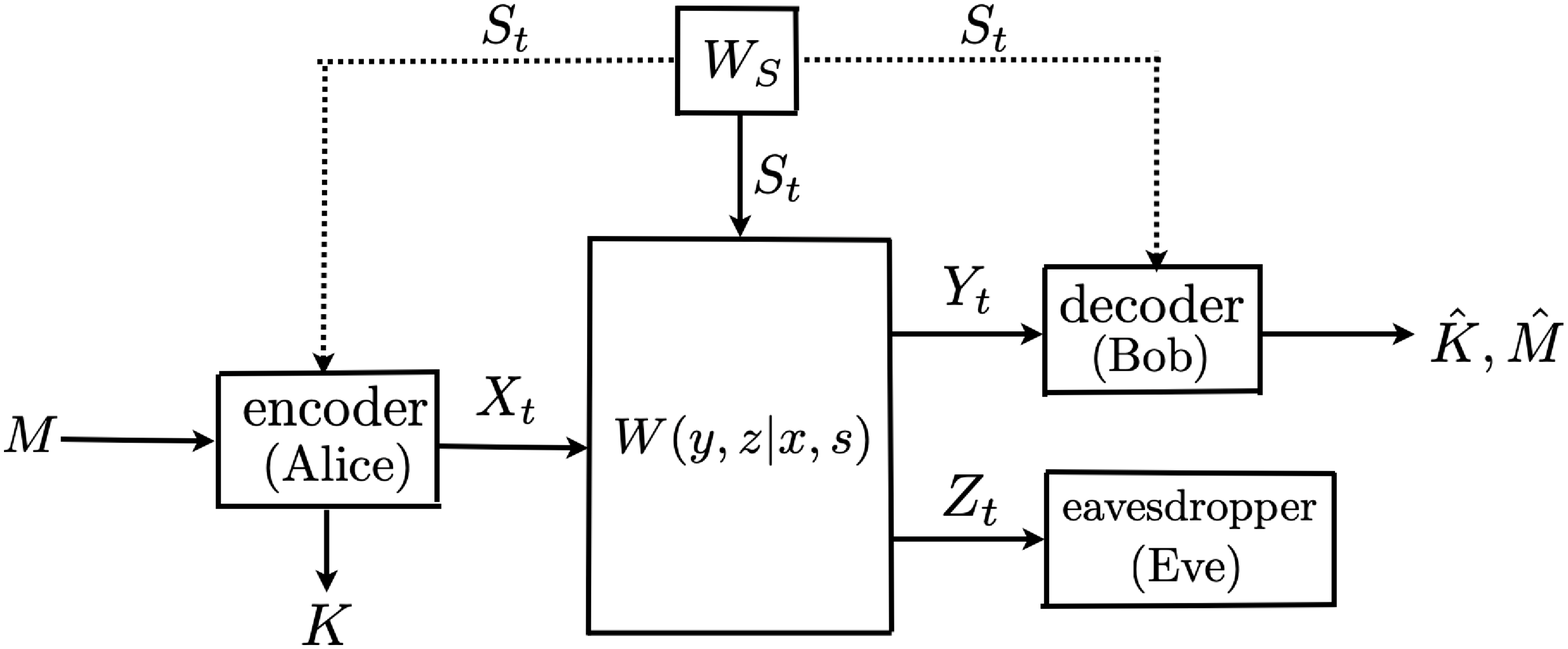}
\end{center}
\caption{WTC with the same CSI available at  Alice and Bob ($t=1,2,\cdots, n$).
          }
\label{fig2}
\end{figure}

In  this section,
we address the problem of converse part (outer bound) for Theorem \ref{teiri:1} (inner bound). Specifically, 
 we establish  the exact  SM-SK capacity region for WTCs with {\em non-causal} CSI
available at ``both" Alice and Bob as in Fig. \ref{fig2}. To do so, 
let the corresponding  non-causal SM-SK capacity region 
be denoted by $\cC_{\mbox{{\scriptsize\rm NCSI-ED}}}$.
\footnote{ED denotes Encoder=E and  Decoder=D.}
%
Moreover, let  $\overline{\cR}_{\sf in}(p_{SUV})$ denote the set of all nonnegative rate pairs $(R_M, R_K)$ satisfying
the rate constraints:
\beqn
R_M &\le& I(UV; Y|S),\label{eq:non2d}\\
R_M+R_K &\le& I(V;Y|SU)-I(V;Z|SU) \non\\
&& \qquad\qquad+H(S|ZU),\label{eq:non3d}
%
\eeqn
where $UV$ may be dependent on $S$, and $H(\cdot), H(\cdot|\cdot)$ denote the (conditional) entropy.
 Then, we have 
\bteiri[Non-causal  SM-SK capacity region]\label{teiri:3}
\beq\label{eq:non-4d}
\cC_{\mbox{{\scriptsize\rm NCSI-ED}}} = \overline{\cR}_{\sf in}
 \stackrel{\Delta}{=} 
  \bigcup_{p_{SUV}} \overline{\cR}_{\sf in}(p_{SUV}),
\eeq
where  the union  is taken over all  ``non-causally" achievable probability distributions. 
Here,
the cardinalities of $U, V$ may be restricted 
to
$|\cU|\le (|\cX|-1) |\cS|+2$ and $ |\cV|\le (|\cX|-1)^2|\cS|^2+2(|\cX|-1)|\cS|+2$,
so that the right-hand side of (\ref{eq:non-4d}) is a compact set.
\IEEEQED
\eteiri

\bchui\label{chui:optimal-1}
{\rm
Theorem \ref{teiri:3}, in particular, means the ``optimality" of the non-causal  inner bound (Theorem \ref{teiri:1}) given by  Bunin {\em et. al}   \cite{bunin-2019} 
when the CSI  $S$ is available at both  Alice and Bob.
}
\IEEEQED
\echui

%
%

%
{\em Proof of achievability for Theorem \ref{teiri:3}:}

The achievabilty immediately follows from Theorem \ref{teiri:1} with $SV, SY$ instead of  
 $V, Y$  in (\ref{eq:non2}) and (\ref{eq:non3}), that is,
 %
 
 %
\beqn
R_M &\le& I(USV; SY)-I(USV;S)\non\\
 & =&  I(USV; SY)-H(S)\non\\ 
 & =&I(UV;Y|S)\label{eq:non2-8};\\
R_M+R_K &\le& I(SV;SY|U)-I(SV;Z|U)\non\\
&& \qquad\qquad - [I(U;S)-I(U;SY)]^+\non\\
&=&  I(SV;SY|U)-I(SV;Z|U)\non\\
& =& I(V;Y|SU) - I(V;Z|SU) \non\\
& &\qquad\qquad+H(S|ZU),
\label{eq:non3-8}
\eeqn
 where we have noticed that 
$I(U;SY) \ge I(U;S)$ and hence $[I(U;S)-I(U;SY)]^+ =0$, and also that $I(USV;S)=H(S).$

\noindent
{\em Proof of converse for Theorem \ref{teiri:3}:}

Suppose that $(R_M, R_K)$ is achievable,
 and set $\overline{Y}^n =S^nY^n$.
 It suffices here  to assume that $M$ is uniformly distributed on $\cM_n$. 

1) We first show (\ref{eq:non2d}).  
Observe that $H(M|\barY^n) \le n\vep_n$ holds by Fano inequality, where $\vep_n \to0$ as $n$ tends to $\infty$.
 Then,
noting that  $S^n$ and $M$ are independent, we have
\beqn
\lefteqn{nR_M}\non\\
 & = & H(M)\non\\
&\le & H(M)-H(M|\barY^n)+n\vep_n\non\\
&=& I(M;\barY^n) +n\vep_n\non\\
&=& I(MS^n; \barY^n) -I(S^n;\barY^n|M)+n\vep_n\non\\
&\le & I(MS^n; \barY^n) -H(S^n|M)+2n\vep_n\non\\
&=& I(MS^n; \barY^n) -H(S^n)+2n\vep_n\non\\
&=& \sum_{t=1}^{n}I(MS^n;\barY_t|\barY^{t-1}) -\sum_{t=1}^n H(S_t )+2n\vep_n\non\\
&\le& \sum_{t=1}^{n}I(MS^n\barY^{t-1};\barY_t) -\sum_{t=1}^n H(S_t )+2n\vep_n\non\\
&\le& \sum_{t=1}^{n}I(MS^n\barY^{t-1}Z_{t+1}^{n};\barY_t) -\sum_{t=1}^n H(S_t )+2n\vep_n\non\\
&\le&  \sum_{t=1}^{n}I(MKS^n\barY^{t-1}Z_{t+1}^{n};\barY_t) -\sum_{t=1}^n H(S_t )+2n\vep_n\non\\
&=& \sum_{t=1}^{n}I(U_tS_tV_t;\barY_t) -\sum_{t=1}^n H(S_t )+2n\vep_n\label{eq:stte1}\\
&=& \sum_{t=1}^{n}I(U_tS_tV_t;S_tY_t) -\sum_{t=1}^n H(S_t )+2n\vep_n,
\label{eq:sune5}
\eeqn
where we have set
\beq\label{eq:sune6}
U_t=\barY^{t-1}Z_{t+1}^{n}, \ V_t =MKS^{t-1}S_{t+1}^{n}.
\eeq
Let us now consider the random variable $J$ such that $\Pr\{J=t\} =1/n \ (t=1,2,\cdots,n).$ 
Then, (\ref{eq:sune5}) is written as
\beqn
R_M &\le & I(U_JS_JV_J; S_JY_J|J) - H(S_J|J ) +2\vep_n\non\\
&\le & I(U_JJS_JV_J; S_JY_J) - H(S_J|J) +2\vep_n\non\\
&= & I(U_JJS_JV_J; S_JY_J) - H(S_J) +2\vep_n\non\\
&= & I(USV; SY) - H(S) +2\vep_n\non\\
&=& I(UV;Y|S) + 2\vep_n,
\label{eq:sune7}
\eeqn
where, noting that $S^n$ is stationary and memoryless and hence $H(S_J|J)=H(S_J)=H(S)$, we have set 
\beq\label{eq:sune8}
U=U_JJ, \ V=V_J, \ S=S_J, \ Y=Y_J, \ Z=Z_J.
\eeq
Thus, by letting $n\to\infty$ in (\ref{eq:sune7}), we obtain (\ref{eq:non2d}).
It is obvious here that $UV\to XS\to YZ$ forms a Markov chain, where we have similarly set $X=X_J$.

2) Next, we show (\ref{eq:non3d}). 
First observe that, in view of Definitions \ref{teigi:3} $\sim$ \ref{teigi:5} in Section \ref{sec-non-causal-WTC} 
as well as 
the uniform continuity of entropy (cf. \cite[Lemma 2.7]{csis-kor-2nd}), we have
\beq
%
|H(K|M=m) - H(U_K)| \le n\varepsilon_n   \mbox{ for all } m \in M_n,\non
\eeq
where $U_K$ denotes the random variable uniformly distributed on  $\cK_n$.
In addition, recall that $M$ is uniformly distributed on $\cM_n$, and therefore
\beqn
nR_M &=& H(M),\non\\
nR_K &=& H(U_K) \le H(K|M=m)  +n\vep_n \mbox{ for all } m \in M_n,\non
\eeqn
which yields 
\beq
nR_M = H(M), \ nR_K \le H(K|M) +n\vep_n.  \nonumber
\eeq
Since $I(MK;Z^n) \le n\vep_n$ by assunption and 
$H(MK|\barY^n) \le n\vep_n$ by Fano inequality, we obtain
\beqn
\lefteqn{n(R_M +R_K)}\non\\
 &\le & H(M)+H(K|M)  +n\vep_n\non\\
&=& H(MK)  +n\vep_n\non\\
& \le & H(MK) -H(MK|\barY^n) +2n\vep_n\non\\
&= & I(MK;\barY^n) +2n\vep_n\non\\
&\le&  I(MK;\barY^n) -I(MK;Z^n)+3n\vep_n.
\label{eq:sune13}
\eeqn
On the other hand, 
\beqn
\lefteqn{I(MK;\barY^n) }\non\\
&=& I(MKS^n;\barY^n)-I(S^n;\barY^n|MK)\non\\
&=&  I(MKS^n;\barY^n)-H(S^n|MK)\non\\
&&  \qquad\qquad+H(S^n|MK\barY^n)
 \label{eq:sune9}
\eeqn
and similarly
\beqn\label{eq:sune11}
\lefteqn{I(MK;Z^n)}\non\\
& = & I(MKS^n;Z^n)-H(S^n|MK) \non\\
&& \qquad\qquad+H(S^n|MKZ^n).
\eeqn
%
Thus, inequality (\ref{eq:sune13}) is continued to 
\beqn
\lefteqn{n(R_M +R_K)}\non\\
 &\le & I(MKS^n;\barY^n) - I(MKS^n; Z^n)\non\\
 & &-H(S^n|MKZ^n) +H(S^n|MK\overline{Y}^n)+3n\vep_n\label{eq:sune10}\\
&\le & I(MKS^n;\barY^n) - I(MKS^n; Z^n) \non\\
&&\qquad\qquad \qquad+H(S^n|MK\overline{Y}^n) +3n\vep_n\label{eq:rev-1}\\
&\le & I(MKS^n;\barY^n) - I(MKS^n; Z^n)  +4n\vep_n\label{eq:sune19}\\
&=&\sum_{t=1}^n I(MKS^n; \barY_t|\barY^{t-1})\non\\
& & \qquad \qquad-\sum_{t=1}^n I(MKS^n; Z_t|Z_{t+1}^{n})+4n\vep_n\non\\
&\stackrel{(c)}{=} & \sum_{t=1}^n I(MKS^nZ_{t+1}^n;\barY_t|\barY^{t-1})\non\\
& & -
\sum_{t=1}^n I(MKS^n\barY^{t-1}; Z_t|Z_{t+1}^{n})+4n\vep_n\non\\
&\stackrel{(d)}{=} & \sum_{t=1}^n I(MKS^n;\barY_t|\barY^{t-1}Z_{t+1}^n)\non\\
& & \qquad \qquad -
\sum_{t=1}^n I(MKS^n; Z_t|\barY^{t-1}Z_{t+1}^{n})+4n\vep_n\non\\
&\stackrel{(e)}{=} &  \sum_{t=1}^n I(S_tV_t;\overline{Y}_t|U_t)-
\sum_{t=1}^n I(S_tV_t; Z_t|U_t)+4n\vep_n,\label{eq:br1}\non\\
&& \label{eq:br1}\\
&=&  \sum_{t=1}^n I(S_tV_t;S_tY_t|U_t)-
\sum_{t=1}^n I(S_tV_t; Z_t|U_t)+4n\vep_n\non\\
&=&
\sum_{t=1}^n I(V_t;Y_t|S_tU_t)-
\sum_{t=1}^n I(V_t; Z_t|S_tU_t)\non\\
&& \qquad\qquad+\sum_{t=1}^nH(S_t|Z_tU_t)+4n\vep_n,
\label{eq:sune15}
\eeqn
where $(c)$ and $(d)$ follow from Csisz\'ar identity (cf. \cite{gamal-kim});
$(e)$  comes from (\ref{eq:sune6}).

Therefore, using (\ref{eq:sune8}), we have
\beq
R_M +R_K \le  I(V;Y|SU)-
 I(V; Z|SU)+H(S|ZU)
 +4\vep_n.\label{eq:sune16}
\eeq
Thus, letting $n\to \infty$ in (\ref{eq:sune16}), we conclude (\ref{eq:non3d}), 
thereby completing the proof of Theorem \ref{teiri:3}.
\IEEEQED

An immediate consequence of Theorem \ref{teiri:3} is  the following two corollaries, where we let
 $C^{{\scriptsize\rm M}}_{\mbox{{\scriptsize\rm NCSI-ED}}}$ (called the SM capacity)
denote
the supremum of the projection of $\cC{\mbox{{\scriptsize\rm NCSI-ED}}}$ on the $R_M$-axis,
and 
 $C^{{\scriptsize\rm K}}_{\mbox{{\scriptsize\rm NCSI-ED}}}$ (called the SK capacity)
denote
the supremum of the projection of $\cC_{\mbox{{\scriptsize\rm NCSI-ED}}}$ on the $R_K$-axis.

Then, we have, with $UV$ and $S$ that may be correlated,
\bkei[Non-causal SM capacity\label{kei:kiku-1}]
{\rm 
\beqn\label{eq:kiku-2}
\lefteqn{C^{{\scriptsize\rm M}}_{\mbox{{\scriptsize\rm NCSI-ED}}}}\non\\
 &=&
\max_{p_{SUV}}\min (I(V;Y|SU)-I(V;Z|SU)+H(S|ZU),\non\\
& & \qquad\qquad\qquad I(UV;Y|S)).
\eeqn
%
}
\ekei
\bkei[Non-causal  SK capacity\label{kei:kiku-2}]
{\rm 
\beqn\label{eq:kiku-3}
\lefteqn{C^{{\scriptsize\rm K}}_{\mbox{{\scriptsize\rm NCSI-ED}}} }\non\\
&=&
\max_{p_{SUV}} (I(V;Y|SU)-I(V;Z|SU)+H(S|ZU)).\non\\
&& 
\eeqn
%
}
\ekei

\bchui\label{chui:kiku-3}
{\rm
The variable $U$ in (\ref{eq:kiku-3}) appears to play the role of  ``time-sharing" parameter, so one may wonder
 if this $U$ can be omitted as in 
 Khisti {\em et al.} \cite[Theorem 3]{khisti-wornell} who
have, instead of (\ref{eq:kiku-3}), given the following formula with the time-sharing parameter $U$ omitted:
\beqn\label{eq:RS2}
 C^{\mbox{\scriptsize\sf K}}_{\mbox{{\scriptsize\rm NCSI-ED}}}
=
 \max_{p_{SV}} \bigl(I(V;Y|S) - I(V;Z|S)+H(S|Z) \bigr).
 \eeqn
 It is evident that the achievability in formula (\ref{eq:kiku-3}) subsumes that of formula (\ref{eq:RS2})
in that  we can set $U=\emptyset$ in (\ref{eq:kiku-3}) to get (\ref{eq:RS2}).
We notice here also  that, as will be seen from the proof of Theorem \ref{teiri:4}, if the WTC in consideration is a degraded one 
($Z$ is a degraded version of $Y$), then
 the right-hand sides  of both (\ref{eq:kiku-3}) and (\ref{eq:RS2}) boil down  to the  right-hand side of (\ref{eq:hop-3}) in
 Corollary \ref{kei:tuika-1}.
  Nevertheless, the $U$ cannot be omitted in general, because
 in maximizing (\ref{eq:kiku-3})
 the ``time-sharing" parameter $U$ cannot necessarily  be
 selected so as to be independent of the  given ``state" $S$ (see ``technical flaws"  in the converse proofs of \cite[Theorem 3]{khisti-wornell}
and \cite[Theorem 1]{khisti-wornell-seoul}).
We are thus tempted to think about the following conjecture:

{\em Conjecture:}
%
{\rm There exists a WTC with non-causal CSI $S$ at both Alice and Bob such that
\beqn\label{pro:3q}
\lefteqn{\max_{p_{SUV}} (I(V;Y|SU)-I(V;Z|SU)+H(S|ZU))}\non\\
&>& 
 \max_{p_{SV}} \bigl(I(V;Y|S) - I(V;Z|S)+H(S|Z) \bigr),
\eeqn
which then means that 
formula (\ref{eq:RS2}) is not  tight in general.
\IEEEQED
}
}
\echui
%

%

\section{ Wiretap channel with Causal CSI}\label{intro-ge-cs1}

%
The encoding scheme in \cite{bunin-2019} used to prove Theorem \ref{teiri:1} is based on 
the  soft covering lemma  as well as  
 the ``non-causal"  likelihood encoding \cite{eva}.
Since the re-interpretation of  this scheme from the ``causal" viewpoint is the very point to be invoked in this section, 
we  here summarize the (non-causal) encoding scheme given by  \cite{bunin-2019}.

{\em Codebook $\cB_n$:}
Define the index sets $\cI_n \stackrel{\Delta}{=}  [1: 2^{nR_1}]$ and ${\cal J}_n \stackrel{\Delta}{=}  [1: 2^{nR_2}]$. 
For each $i\in \cI_n $, generate $\ssu_i \in \cU^n$ of length $n$ that are i.i.d. according to probability distribution
\footnote{$p_U^n$ for a random variable $U$ denotes the $n$ times product probability distribution of $p_U$.
Similarly for $p_{V|U}^n$.} 
$p_U^n$. Next, given   $i\in \cI_n $, for each $(j, k, m) \in {\cal J}_n\times {\cal K}_n \times {\cal M}_n$
generate $\ssv_{ijkm} \in \cV^n$ that are  i.i.d. according to 
conditional probability distribution $p^n_{V|U}(\cdot | \ssu_i).$

{\em Likelihood encoder $f_n$:}
 Given $m\in \cM_n$ and $\sss\in \cS^n$, the encoder ``randomly" chooses 
 $(i, j, k) \in {\cal I}_n\times {\cal J}_n \times {\cal K}_n$ according to the conditional probability ratio ``proportional" to
 \beq\label{eq:non-s1}
f_{\sf LE}(i,j,k|m, \sss)   \stackrel{\Delta}{=}  p^n_{S|UV}(\sss|\ssu_i, \ssv_{ijkm}),
 \eeq
where $p_{S|UV}$ is the conditional probability distribution induced from $p_{SUVX}$.
The encoder declares the chosen index $k\in \cK_n$ as the key.  Given the chosen $(\ssu_i, \ssv_{ijkm})$,
the channel input sequence $\ssx\in \cX^n$ is generated according to conditional probability distribution
$p_{X|SUV}^n(\cdot|\sss, \ssu_i, \ssv_{ijkm})$.

{\em Decoder $\phi_n$:} Upon observing the channel output $\ssy\in\cY^n$, the decoder searches for a unique
$(\hat{i},\hat{j},\hat{k}, \hat{m})$ $ \in {\cal I}_n\times{\cal J}_n\times {\cal K}_n \times {\cal M}_n$
such that 
\beq\label{eq:non-s2}
(\ssu_{\hat{i}}, \ssv_{\hat{i}\hat{j}\hat{k}\hat{m}}, \ssy)\in \cT_{\epsilon}^n(p_{UVY}),
\eeq
where $\cT_{\epsilon}^n(p_{UVY})$ denotes the  set of jointly $\vep$-typical sequences
(cf. \cite{csis-kor-2nd}).
If such a unique quadruple is found, then set $\phi_n(\ssy) = (\hat{m}, \hat{k})$.
Otherwise, $\phi_n(\ssy) = (1, 1)$.
\bchui\label{chui:joint-likelihood}
{\rm
Roughly speaking, the likelihood encoder $f_n$ can be regarded  as a
 {\em smoothed} version of the joint typicality encoder 
(cf. Gelfand and Pinsker \cite{gel-pin})  that,
given $\sss$, picks up ``at random" sequences $(\ssu_i, \ssv_{ijkm})$
with larger weights on jointly  typical  (with  $\sss)$ sequences  and smaller weights on jointly atypical sequences.
\IEEEQED
}
\echui

Theorem \ref{teiri:1}  is of crucial significance in the sense that this provides 
 the ``best"  inner bound 
to subsume,  in a unifying way,  all the known results in this field
 for WTCs with  ``{\em non-causal}" CSI available at Alice.
 As such, on the other hand, at first glance Theorem \ref{teiri:1} does not appear to give any insights into WTCs with 
 ``{\em causal}" CSI. However, for the region $\cR_{\sf in}(p_{SUV})$  with a class of some simple but relevant $UV$s, 
 it is possible to re-interpret $\cR_{\sf in}(p_{SUV})$ as inner bounds for WTCs with  ``{\em causal}" CSI at Alice.
 This operation is called  {\em plugging},
which is  developed hereafter.

The ``causal  code" that we consider in this section is the following, which is the causal counterpart of 
 the non-causal code defined as in Definition \ref{teigi:1}:
 \bteigi[Causal code]\label{teigi:causal-1}
 %
{\rm
An $(n, R_M, R_K)$-code $c_n$ for the WTC with ``causal"  CSI  at Alice and  message set 
$\cM_n$ and  key set $\cK_n$ is
a triple of functions $(f_n^{(1)}, f_n^{(2)}, \phi_n)$ such that

 1) $ f_n^{(1)}: \cM_n\times \cS^{t} \to \cP (\cX)  \quad (t=1,2,\cdots, n)$;
 
 2) $f_n^{(2)}: \cM_n\times \cS^n \to {\cal P}(\cK_n)$,
 
 3) $\phi_n: \cY^n \to \cM_n\times \cK_n$, \\
 where  $ f_n^{(1)}, f_n^{(2)}$ are stochastic functions.
 }
\IEEEQED
  \eteigi
  \bchui\label{chui:tesuya1}
  {\rm
  One may wonder if $f_n^{(2)}$ in the above should be $f_n^{(2)}: \cM_n \to\cK_n$
 because we are here considering ``causal" encoders but $f_n^{(2)}$ here looks to require $S^n$ at once 
 before the beginning of encoding at Alice.
 However, actually, the operation $f_n^{(2)}: \cM_n\times \cS^n \to  \cP (\cK_n) $ can be carried out by Alice at the end of
  the current block (of length $n$).
  This is possible with causal codes.
  \IEEEQED
  }
  \echui
%
%
\bteigi[Causal SM-SK capacity region]\label{teigi:8}
{\rm
%
The SM-SK capacity region  of  the WTC with 
``causal"  CSI at Alice,
denoted by $\cC_{\mbox{{\scriptsize\rm CSI-E}}}$, 
 is the 
 set of all causally  SM-SK achievable rate pairs with CSI at Alice, and 
the supremum of the projection of $\cC_{\mbox{{\scriptsize\rm CSI-E}}}$ on the $R_M$-axis, denoted by
 $C^{{\scriptsize\rm M}}_{\mbox{{\scriptsize\rm CSI-E}}}$,
 is called the SM capacity,
whereas 
the supremum of the projection of $\cC_{\mbox{{\scriptsize\rm CSI-E}}}$ on the $R_K$-axis is called the SK capacity,
denoted by
 $C^{{\scriptsize\rm K}}_{\mbox{{\scriptsize\rm CSI-E}}}$.
 %
 \IEEEQED
} 
\eteigi

\bteigi[Causal  achievability]
\label{sinnen}
{\rm
We now consider the following special class  of random variables $UV$'s 
such that there exists  some $\tilde{U}\tilde{V}$ independent of $S$ 
 ($\tilde{U}$ and $\tilde{V}$ may be correlated)
 for which
 \beqn
& Case\ 1):   & V=\tilde{V},\ U=\tilde{U}; \label{eq:non-3s}\\
&  Case\ 2):   &V= (S, \tilde{V}),\  U= \tilde{U}; \label{eq:non-4s}\\
 & Case\ 3):  &  V=\tilde{V}, \ U= (S, \tilde{U}); \label{eq:non-5s}\\
 &Case\ 4):  &  V= (S, \tilde{V}),\ U= (S, \tilde{U}). \label{eq:non-6s}
 \eeqn
We say  that the probability distribution $p_{YZSXUV}$ (or the corresponding random variable 
$YZSXUV$) is {\em causally achievable}  if, in addition to (\ref{eq:non1})
and the independence of $S$ and $\tU\tV$, 
 one of conditions (\ref{eq:non-3s}) $\sim$ (\ref{eq:non-6s})  is satisfied.
%
%
\IEEEQED
}
\eteigi

\begin{figure}[htbp]
\begin{center}
\includegraphics[width=50mm]{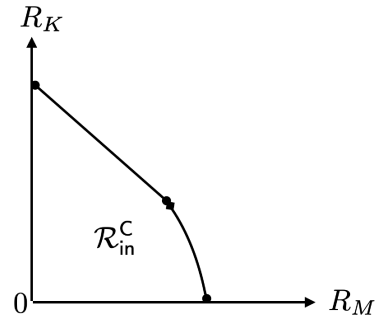}
\end{center}
\caption{Causal SM-SK achievable rate region.
          }
\label{fig0}
\end{figure}

With these preparations, we have the following causal version of Theorem \ref{teiri:1} (cf. Fig. \ref{fig0}),
where 
$\cR_{\sf in}^{{\sf N}}$
as in Section \ref{sec-non-causal-WTC}
is replaced here by 
the causally achievable region $
\cR_{\sf in}^{\sf C}
$.
%
%
%
\bteiri[Causal  SM-SK inner bound]\label{teiri:2}
\beq\label{eq:non-s10}
\cC_{\mbox{{\scriptsize\rm CSI-E}}} \supset \cR_{\sf in}^{\sf C}
 \stackrel{\Delta}{=}
  \bigcup_{{\sf C:}p_{SUV}} \cR_{\sf in}(p_{SUV}),
\eeq
where  the union is taken over all ``causally" achievable probability distributions $p_{SUV}$'s
and $\cR_{\sf in}(p_{SUV})$ is  the same one as  in Theorem \ref{teiri:1}.
\IEEEQED
\eteiri

 {\em Proof:}
  In this proof too, under all Definitions \ref{teigi:1} $\sim$ \ref{teigi:5} with Definition \ref{teigi:1}
 replaced by Definition \ref{teigi:causal-1}, we invoke the same Codebook $\cB_n$ and the likelihood encoder 
 $f_n$  
 as  in Section \ref{sec-non-causal-WTC}. The point here is to show that the likelihood encoder  $f_n$ can in fact be 
 implemented in a causal way for causally achievable probability distributions $p_{SUV}$'s.

Although it may look to be necessary to give the proofs for each of Case 1)  $\sim$  Case 4),
the ways of those proofs are essentially the same, so it suffices,  without loss of generality, 
 to show that the likelihood encoder $f_n$ 
can actually be  implemented for Case 2) in a causal way.

First, recall that, in Case 2), $p_{S|UV}\equiv p_{S|US\tilde{V}}$ is the conditional distribution of $S$ given 
$UV=US\tilde{V}$
and hence, irrespective of  $u, \tilde{v}$,
\beq\label{eq:bn1}
p_{S|US\tilde{V}}(s|  u, s^{\prime},\tilde{v})
= \left\{
\begin{array}{cl}
1 & \mbox{if}  \ s=s^{\prime}, \\
0 & \mbox{if}  \ s\neq s^{\prime}.
\end{array}
\right.
\eeq
 Then, since $p^n$ is a product probability distribution (i.e., memoryless) of $p$,
setting as $\ssv_{ijkm} = (\sss_{ijkm}, \tilde{\ssv}_{ijkm})$, the conditional probability ratio  in 
 (\ref{eq:non-s1}) can be evaluated as follows.
\beqn\label{eq:memo1}
\lefteqn{f_{\sf LE}(i,j,k|m, \sss)}\non\\
 &=& 
 p^n_{S|UV}(\sss|\ssu_i, \ssv_{ijkm})\non\\
  &=& p^n_{S|US\tilde{V}}(\sss|\ssu_i, \sss_{ijkm}, \tilde{\ssv}_{ijkm})\non\\
 &=& \prod_{t=1}^{n}p_{S|US\tilde{V}}(s^{(t)}| u_i^{(t)}, s^{(t)}_{ijkm},\tilde{v}^{(t)}_{ijkm}), %
 \eeqn
where  we have set 
\beqn
\sss &=& (s^{(1)}, s^{(2)}, \cdots, s^{(n)}),\label{eq:ko-1}\\
\ssu_i &=& (u_i^{(1)}, u_i^{(2)}, \cdots, u_i^{(n)}),\label{eq:ko-2}\\
\sss_{ijkm} &=&  (s_{ijkm}^{(1)}, s_{ijkm}^{(2)}, \cdots, s_{ijkm}^{(n)}),\label{eq:ko-3}\\
\tilde{\ssv}_{ijkm} &=&  (\tilde{v}_{ijkm}^{(1)}, \tilde{v}_{ijkm}^{(2)} \cdots, \tilde{v}_{ijkm}^{(n)})\label{eq:ko-4}.
\eeqn
%

Now, in view of  (\ref{eq:bn1}), it turns out that $p_{S|US\tilde{V}}(s^{(t)}| u_i^{(t)}, s^{(t)}_{ijkm},\tilde{v}^{(t)}_{ijkm})$ in 
(\ref{eq:memo1}) is equal to 1 if $s^{(t)} = s^{(t)}_{ijkm}$; otherwise, equal to  0 ($t=1,2,\cdots, n)$, so that we have,
%
irrespective of $(\ssu, \tilde{\ssv})$,
\beq\label{eq:bn2}
p^n_{S|US\tilde{V}}(\sss|\ssu, \sss_{ijkm}, \tilde{\ssv})= \left\{
\begin{array}{cl}
1 & \mbox{if}  \ \sss_{ijkm}=\sss, \\
0 & \mbox{if}  \  \sss_{ijkm}\neq \sss.
\end{array}
\right.
\eeq
Therefore, in particular,
%
\beq\label{eq:saiki-1}
p^n_{S|US\tilde{V}}(\sss|\ssu_{i}, \sss, \tilde{\ssv}_{ijkm})=1 \ \mbox{for all}\ 
(i,j,k)\in\cI_n\times {\cal J}_n\times \cK_n,
\eeq
so that, 
 given $(m,\sss)$,  the stochastic (non-causal) likelihood encoder $f_n$ as specified 
 in Section \ref{sec-non-causal-WTC}
 chooses $(\ssu_{i}, \sss, \tilde{\ssv}_{ijkm})$ {\em uniformly} over the set
\beq\label{eq:suhyan-1}
\cL(m, \sss)\stackrel{\Delta}{=} \{(\ssu_{i}, \sss,  \tilde{\ssv}_{ijkm})|
(i,j,k)\in\cI_n\times {\cal J}_n\times \cK_n \}.
\eeq
We notice here that, since
$U\tilde{V}$ and $S$ are independent and hence $(\ssu_{i},  \tilde{\ssv}_{ijkm})$,
$\sss_{ijkm}$  and $\sss$ are also mutually independent,
  the set 
  \beq\label{eq:sanko-2}
 \cL(m) 
 \stackrel{\Delta}{=} \{(\ssu_{i},  \tilde{\ssv}_{ijkm})|
(i,j,k)\in\cI_n\times {\cal J}_n\times \cK_n \}
 \eeq
can actually be generated in advance of encoding, {\em not} depending on $(\sss_{ijkm}, \, \sss).$

%
 Up to here, it was assumed that the full state information $\sss$
 is non-causally available at the encoder, so
the point  here is how this non-causal encoder $f_n$ 
can be replaced by a causal encoder. 
%
This is 
indeed possible, because 
 $\sss_{ijkm}= \sss$ can be written  componentwise as $s^{(t)}_{ijkm} =s^{(t)}\ (t=1,2,\cdots, n)$ and therefore
the encoder can set $s^{(t)}_{ijkm}$ to be $s^{(t)}$ at each time $t$ using the state information 
$s^{(t)}$
available  at time $t$ at the encoder,
which clearly can be  carried out in the ``causal" way. Moreover, 
  $(\ssu_i,  \tilde{\ssv}_{ijkm})$  can also be fed 
 in  the causal way (componentwise) according as 
 $(u_i^{(t)},\tilde{v}_{ijkm}^{(t)})$ ($t=1,2,\cdots, n)$, because  $(\ssu_i,  \tilde{\ssv}_{ijkm})$ 
  was generated in advance of encoding.

%
Thus,
given the chosen $(\ssu_i, \sss, \tilde{\ssv}_{ijkm})$, the encoder generates
the channel input sequence 
$$\ssx=(x^{(1)}, x^{(2)}, \cdots, x^{(n)})  \in \cX^n$$
 according to the conditional probability:
\beqn\label{eq:sukki-12p}
\lefteqn{p_{X|SUS\tilde{V}}^n(\ssx|\sss,  \ssu_i, \sss, \tilde{\ssv}_{ijkm})}\non\\
 &=&
 \prod_{t=1}^{n}p_{X|SUS\tilde{V}}(x^{(t)}|  s^{(t)}, u_i^{(t)}, s^{(t)},\tilde{v}^{(t)}_{ijkm}),
\eeqn 
which implies that the $\ssx$ can also be generated   in the causal way
according as  $x^{(t)}\  (t=1,2, \cdots, n)$,
thereby completing the proof of Theorem \ref{teiri:2}.
\IEEEQED
\smallskip

So far in this section we have invoked, as a crucial step,  the argument of plugging, 
the logical core of which is schematically summarized  as follows:
\bmeidai[Principle of plugging]\label{prop:1a}
{\rm
Consider a channel coding system  (memoryless but not necessarily WTCs)
with CSI $S$ and auxiliary random variables 
$U_1, U_2, \cdots, U_a$ together with rate tuple $(R_1, R_2, \cdots, R_b)$
to be used for generation of the  random code
\beqn\label{eq:sakura-q1}
 \lefteqn{\cC=}\non \\
 & &
 \left\{
(\ssu_{1i_1},\ssu_{2i_2},\cdots, \ssu_{ai_a})
\right\}_{i_1\in[1:2^{nR^{\prime}_{1}}], i_2\in[1:2^{nR^{\prime}_{2}}], \cdots, i_a\in[1:2^{nR^{\prime}_{a}}]
}
\non\\
&&
\eeqn
where
each $R^{\prime}_k$ ($k=1,2,\cdots, a$) ia a partial sum of $R_1,R_2,\cdots, R_b$
 (for example, $R^{\prime}_1=R_1+R_3, R^{\prime}_2=R_2$, etc.) and 
 each codeword
 $(\ssu_{1 i_1},\ssu_{2i_2},\cdots, \ssu_{ai_a})$
  is generated 
 according to product probability distribution $p^n_{U_1U_2\cdots U_a}$ (or its marginal conditional distributions).
Given message
 $m$ and state sequence $\sss$,
the non-causal  (likelihood) encoder
 $f_n$  stochastically
picks \footnote{$f_n$ may  also be a joint typicality encoder (cf. Example \ref{chui:perp}).}
%
an element of $\cC$
and maps it ``componentwise"  to a channel input $\ssx$ according to 
conditional probability distribution 
 $p^n_{X|SU_1U_2 \cdots  U_a}(\cdot |\sss,\ssu_{1i_1},\ssu_{2i_2},\cdots, \ssu_{ai_a})$. 
Now suppose that 
  any rate tuple $(R_1, R_2, \cdots, R_b)$ satisfying  the rate constraints
\beqn\label{sle-1}
F_1(R_1, R_2,\cdots, R_b; U_1, U_2,\cdots, U_a; S) & \ge & 0,\label{epq:ll1}\\
F_2(R_1, R_2,\cdots, R_b; U_1, U_2,\cdots, U_a; S) & \ge & 0,\label{epq:ll2}\\\
\cdots\cdots\cdots \cdots\cdots\cdots& & \non\\
F_c(R_1, R_2,\cdots, R_b; U_1, U_2,\cdots, U_a; S) & \ge & 0\label{epq:lla}
\eeqn
is ``non-causally"  SM-SK achievable. Then, 
 any rate tuple $(R_1, R_2, \cdots, R_b)$ satisfying  the rate constraints (\ref{epq:ll1}) $\sim$ (\ref{epq:lla}) with
 \beqn\label{eq:finish-1}
U_1 &=& \tU_1 \mbox{\ or\ } (S,\tU_1);  U_2 = \tU_2  \mbox{\ or\ } (S, \tU_2);\non\\
& &\qquad\qquad \qquad\cdots ;  U_a = \tU_a  \mbox{\ or\ } (S,\tU_a)
\eeqn
is ``causally"  SM-SK achievable, where $\tU_1, \tU_2, \cdots, \tU_a$ 
(may be correlated) are independent of $S$. 
}
\IEEEQED
\emeidai
%


\brei\label{chui:perp}
{\rm
A simple  example (with $Z \equiv \emptyset$ (constant variable))
is the relation of the Gelfand-Pinsker (non-causal) coding \cite{gel-pin} and the Shannon strategy (causal) coding
\cite{shannon-cent}.
The former gives the formula 
\beq\label{peq:asa-hi-1}
C^{\rm M}_{\mbox{{\scriptsize\rm NCSI-E}}}  = \max_{p_{SU}}(I(U;Y)-I(U;S)),
\eeq
while the latter gives the formula
\beq\label{peq:asa-hi-2}
C^{\rm M}_{\mbox{{\scriptsize\rm CSI-E}}}  = \max_{p_Sp_U}I(U;Y).
\eeq
Principle of plugging applied to (\ref{peq:asa-hi-1}) claims that, given independent $S$ and $\tilde{U}$, 
rates $R^{\prime}=I(\tU; Y)-I(\tU;S)=I(\tU; Y)$ 
and $R^{\prime\prime}=I(\tU S; Y)-I(\tU S; S)$ $=I(\tU S; Y)-H(S)$ are ``causally" achievable.
It is easy to check that $R^{\prime} \ge R^{\prime\prime}$, so in this case $R^{\prime\prime}$  is redundant.
Thus, the achievablity part of (\ref{peq:asa-hi-2}) is concluded from that of  (\ref{peq:asa-hi-1}) without a separate proof.
\IEEEQED
}
\erei

\section{Applications of Theorem \ref{teiri:2}}\label{sec-application-causal-WTC}

Having established Theorem \ref{teiri:2} on WTCs with causal CSI at Alice, 
in this section we develop it for each of Case 1) $\sim$ Case 4) to demonstrate that, via Theorem \ref{teiri:2}, 
we can unifyingly derive  the previously known  {\em causal} ``lower" bounds 
such as  in 
 \cite{chia-elgamal}, \cite{fujita} and  \cite{han-sasaki-c}.
In addition, we also  demonstrate  that 
 a {\em new}  class of  causal ``inner" bounds directly follow from Theorem \ref{teiri:2},
 which could not have been easily obtained without Theorem \ref{teiri:2}.
 They are largely classified into Propositions \ref{prop:1} and \ref{prop:2}.
In particular, we emphasize that  in this section we are concerned solely with 
 ``two-dimensional"  inner/outer bounds of causally achievable rate pairs $(R_M, R_K)$,
 which
  are derived  in this paper for the first time. 
  %

%
V.A: {\em Causal inner bounds:}

Let us now scrutinize the claim of  Theorem \ref{teiri:2}.
For  the  convenience of discussion, 
we record again here the rate constraints  (\ref{eq:non2}) and (\ref{eq:non3}) as  %
\beqn
R_M &\le& I(UV; Y)-I(UV; S),\label{eq:non-p2}\\
R_M+R_K &\le& I(V;Y|U)-I(V;Z|U) \non\\
& & \qquad- [I(U;S)-I(U;Y)]^+,
\label{eq:non-p33}
\eeqn
which is specifically developed according to Cases 1) $\sim$ 4) as follows.

\smallskip

{\em  Case 1)} : Since $U=\tilde{U}, V=\tilde{V}$ and $\tilde{U}\tilde{V}$ is independent of $S$, 
(\ref{eq:non-p2}) and (\ref{eq:non-p33}) reduce to
\beqn
R_M &\le& I(\tilde{U}\tilde{V}; Y),\label{eq:non-p5}\\
R_M+R_K &\le& I(\tilde{V};Y|\tilde{U})-I(\tilde{V};Z|\tilde{U}),
\label{eq:non-p6}
\eeqn
where we have used $I(\tilde{U}\tilde{V}; S) =0$ and $[I(\tilde{U};S)-I(\tilde{U};Y)]^+=0$.
Clearly, (\ref{eq:non-p5}) is redundant, so only (\ref{eq:non-p6}) remains. Hence,
removing tilde $\tilde{}$ to make the notation simpler, we have
\beq\label{eq:non-p7}
R_M+R_K \le  I(V;Y|U)-I(V;Z|U).
\eeq
It is not difficult to check that replacing (\ref{eq:non-p7}) by
\beq\label{eq:non-p8}
R_M+R_K \le  I(V;Y)-I(V;Z)
\eeq
does not affect the inner region. Thus,
%
%
 \beq\label{tuika-june-R}
\cC_{\mbox{{\scriptsize\rm CSI-E}}} \supset \bigcup_{p_{S}p_{V}}
\{\mbox{rate pairs $(R_M, R_K)$ satisfying
(\ref{eq:non-p8})}\},
\eeq
which implies, in particular,  the  non-causal SM achievability ($R_K=0$) of  Dai and Luo \cite[Theorem 3]{dai-luo}.
%

\smallskip

{\em Case 2)} : Since $U=\tilde{U}, V=S\tilde{V}$ and $\tilde{U}\tilde{V}$ is independent of $S$, 
(\ref{eq:non-p2}) and (\ref{eq:non-p33}) are computed as 
\beqn
R_M &\le&I(\tU S\tV;Y)-I(\tU S\tV;S)\non\\
&=& I(\tU S\tV;Y)-H(S);\label{eq:keio-1}\\
R_M+R_K &\le& I(S\tV;Y|\tU)-I(S\tV;Z|\tU)\non\\
& & -[I(\tU;S)-I(\tU;Y)]^+\non\\
&\stackrel{(a)}{=}&
 I(S\tV;Y|\tU)-I(S\tV;Z|\tU),\label{eq:keio-2}
\eeqn
where $(a)$ follows from $I(\tU;S)=0.$ Therefore, removing tilde $\tilde{}$
again to make the notation simpler, we have the rate constraints for Case 2),
\beqn\label{eq:non-p10}
 R_M &\le & I(U S V;Y)-H(S);\label{eq:keio-3}\\
R_M+R_K &\le&  I(SV;Y|U)-I(SV;Z|U),\label{eq:keio-4-1}
\eeqn
where $UV$ and $S$ are independent. 
Therefore, any nonnegative rate pair $(R_M, R_K)$ is achievable if rate constraints
 (\ref{eq:keio-3}) and (\ref{eq:keio-4-1}) are satisfied.
 Thus, we have the following fundamental inner bound:
\bmeidai[Causal SM-SK inner bound: type I]\label{prop:1}
 \beq\label{tuika-june-O}
\cC_{\mbox{{\scriptsize\rm CSI-E}}} \supset \bigcup_{p_Sp_{UV}}
\{{\mbox{rate pairs $(R_M, R_K)$ satisfying
(\ref{eq:non-p10}) and\ } (\ref{eq:keio-4-1})}\}.
\eeq
\emeidai

An immediate by-product of (\ref{tuika-june-O})  is the following 
corollary:

\bkei[Causal lower bound  (1) at Alice]\label{kei-poiu-1}
{\rm 
\beqn
C^{{\scriptsize\sf M}}_{\mbox{{\scriptsize\rm CSI-E}}}
&\ge & \max_{p_Sp_{UV}}\min ( I(SV;Y|U)-I(SV;Z|U), \non\\
& & \qquad\qquad\qquad I(U SV;Y)-H(S)),\label{eq:silk-1}\\
C^{{\scriptsize\sf K}}_{\mbox{{\scriptsize\rm CSI-E}}} 
&\ge & \max_{\stackrel{p_Sp_{UV}:}{I(US V;Y)\ge H(S)}}
 (I(SV;Y|U)-I(SV;Z|U)),\non\\
 & &\label{eq:silk-2}
\eeqn
where $UV$ and $S$ are independent.
\IEEEQED
}
\ekei
{\em Proof:}
Setting $R_K=0$ in (\ref{tuika-june-O})   yields (\ref{eq:silk-1}),
while setting $R_M=0$  in (\ref{tuika-june-O})   yields (\ref{eq:silk-2}). \IEEEQED
%

Let us now consider two special cases of (\ref{tuika-june-O}).
\smallskip

{\em A:} Let $U=\emptyset$ (constant variable), then (\ref{eq:keio-3}) and (\ref{eq:keio-4-1}) reduce to
\beqn
 R_M &\le & I(SV;Y)-H(S);\label{eq:keio-5}\\
R_M+R_K &\le&  I(SV;Y)-I(SV;Z)\label{eq:keio-6}
\eeqn
with independent $V$ and $S$. 
Consequently, any nonnegative rate pair $(R_M, R_K)$  is achievable if rate constraints 
(\ref{eq:keio-5}) and (\ref{eq:keio-6}) are
satisfied.
Thus, we have
\beq\label{tuika-june-A}
\cC_{\mbox{{\scriptsize\rm CSI-E}}} \supset \bigcup_{p_Sp_{V}}\{{\mbox{rate pairs $(R_M, R_K)$ satisfying
(\ref{eq:keio-5}) and\ } (\ref{eq:keio-6})\}}.
\eeq
%
%
  %

\bchui\label{kei:teyon-1}
{\rm
Setting $R_K=0$ in (\ref{tuika-june-A})  yields the SM lower bound: 
\beq
C^{{\scriptsize\sf M}}_{\mbox{{\scriptsize\rm CSI-E}}} 
\ge  \max_{p_Sp_V}\min ( I(SV;Y)-I(SV;Z), I(SV;Y)-H(S)).\label{eq:silk-10}
\eeq
On the other hand, setting $R_M=0$ in (\ref{tuika-june-A}) yields the SK  lower bound:
\beq
C^{{\scriptsize\sf K}}_{\mbox{{\scriptsize\rm CSI-E}}} 
\ge \max_{\stackrel{p_Sp_{V}:}{I(S V;Y)\ge H(S)}}
 (I(SV;Y)-I(SV;Z)),\label{eq:silk-20}
 \eeq
 which was leveraged, without the proof,  in Han and Sasaki \cite[Remark 5]{han-sasaki-c}.
}
%

Next, in order to compare formula (\ref{eq:silk-10}) with the previous result, 
we develop it in the sequel.
%
 First, (\ref{eq:keio-5}) is rewritten as 
\beqn\label{eq:non-p13}
R_M &\le & I(SV;Y)-H(S)\non\\
 &=& I(V;Y) +I(S;Y|V)-H(S)\non\\
 &\stackrel{(b)}{=}&  I(V;Y) -H(S|VY),
 \eeqn
 where $(b)$ follows from the independence of $V$ and $S$.
 On the other hand, (\ref{eq:keio-6}) is evaluated as follows:
\beqn
\lefteqn{R_M+R_K}\non\\
 &\le& 
I(SV;Y)-I(SV;Z)\non\\  
&=& I(V;Y)+I(S;Y|V)-I(S;Z)-I(V;Z|S)\non\\
&=& I(V;Y)+H(S|V)-H(S|VY) -H(S)\non\\
& & +H(S|Z) -I(V;Z|S)\non\\
&=& I(V;Y)-I(V;SZ) +I(V;S)+H(S|V)\non\\
& & -H(S|VY)-H(S) +H(S|Z)\non\\
&=& I(V;Y)-I(V;SZ)+H(S|Z)-H(S|VY). \label{eq:keio-7}
\eeqn

Summarizing, we have,  with independent $V$ and $S$,
\beqn
R_M  &\le &  I(V;Y) -H(S|VY),\label{eq:non-p13}\\
R_M+R_K &\le& I(V;Y)-I(V;SZ)\non\\
& & \qquad+H(S|Z)-H(S|VY).
 \label{eq:keio-7}
\eeqn
Thus, 
\beq\label{tuika-june-G}
\cC_{\mbox{{\scriptsize\rm CSI-E}}} \supset \bigcup_{p_Sp_{V}}
\{{\mbox{rate pairs $(R_M, R_K)$ satisfying
(\ref{eq:non-p13}) and\ } (\ref{eq:keio-7})\}},
\eeq
which is equivalent to (\ref{tuika-june-A}).
Now, setting $R_K=0$ in (\ref{tuika-june-G}), 
 it turns out that formula (\ref{eq:silk-10})  is rewritten as
  \beqn\label{hasa-1}
    C^{{\scriptsize\sf M}}_{\mbox{{\scriptsize\rm CSI-E}}}
    & \ge & 
    \max_{p_Sp_V}\min(I(V;Y)-I(V;SZ)\non\\
    & &\quad+H(S|Z)-H(S|VY),\non\\
    & & \qquad\qquad I(V;Y) -H(S|VY))
  \eeqn
with independent $V$ and $S$, which was given as
$R_{\mbox{{\scriptsize\rm CSI-1}}}$
by  Han and Sasaki \cite[Theorem 1]{han-sasaki-c} (also cf. Fujita \cite[Lemma 1]{fujita}).
%
%
\IEEEQED
%
\echui
 
 {\em B:} Let $V=\emptyset$, then (\ref{eq:keio-3}) and (\ref{eq:keio-4-1}) reduce to 
 \beqn
 R_M &\le & I(US;Y)-H(S),\label{eq:hop-2a}\\
%
R_M+R_K &\le&  I(S;Y|U)-I(S;Z|U)\label{eq:hop-3b}
\eeqn
with independent $U$ and $S$.
It is easy to check that (\ref{eq:hop-2a}) and (\ref{eq:hop-3b}) are rewritten equivalently as 
\beqn
R_M &\le&  I(U;Y)-H(S|UY), \label{eq:keio-10}\\
R_M+R_K &\le& H(S|UZ)-H(S|UY).\label{eq:keio-10-2}
\eeqn
Consequently, any nonnegative  pair $(R_M, R_K)$  is achievable if  constraints 
 (\ref{eq:keio-10}) and (\ref{eq:keio-10-2}) are
satisfied. Thus,
\beq\label{tuika-june-B}
\cC_{\mbox{{\scriptsize\rm CSI-E}}} \supset \bigcup_{p_Sp_{U}}
\{{\mbox{rate pairs $(R_M, R_K)$ satisfying
(\ref{eq:keio-10}) and\ } (\ref{eq:keio-10-2})\}}.
\eeq
\bchui\label{chui:sune3}
{\rm
Setting $R_K=0$ in  (\ref{tuika-june-B})
 yields the lower bound with independent $U$ and $S$:
  \beqn\label{hasa-2}
C^{{\scriptsize\sf M}}_{\mbox{{\scriptsize\rm CSI-E}}} 
&\ge & \max_{p_Sp_U}
\min(H(S|UZ)-H(S|UY), \non\\
& & \qquad\qquad\qquad I(U;Y)-H(S|UY))
  \eeqn
which was given as s
$R_{\mbox{{\scriptsize\rm CSI-2}}}$
by  Han and Sasaki \cite[Theorem 1]{han-sasaki-c}. 

On the other hand,  setting $R_M=0$ in (\ref{tuika-june-B}),       
we have, for independent $U$ and $S$,
\beqn\label{eq:riri7}
C^{{\scriptsize\sf K}}_{\mbox{{\scriptsize\rm CSI-E}}} \ge \max_{\stackrel{p_Sp_U:}{I(U;Y)\ge H(S|UY)}}(H(S|UZ)-H(S|UY)),
\eeqn
which is a new type of lower bound.
We notice here  that either   (\ref{eq:silk-10}) or (\ref{hasa-2}) does not   always outperform the other. 
  Similarly, we can check that
  either (\ref{eq:silk-20}) or (\ref{eq:riri7})  does not    always outperform  the other. 
  The proof of them is given in  Appendix \ref{addB}.
}
\IEEEQED
\echui

We now have the following two corollaries for WTCs with causal CSI available at ``both" Alice and Bob.
 \bkei[Causal inner bound  (2) at Alice and Bob]\label{kei:chia-el1}
 {\rm 
 Let us consider the WTC with causal CSI  at both Alice and Bob, as depicted in Fig. \ref{fig2}.
 Then, a pair $(R_M, R_K)$ is achievable if the following rate constraints are satisfied:
 \beqn
 R_M &\le & I(V;Y|S);\label{eq:keio-12}\\
R_M+R_K &\le&  I(V;Y|S)-I(V;Z|S) \non\\
& &\qquad+H(S|Z),\label{eq:keio-13-2}
\eeqn
where $V$ and $S$ are independent.
Thus, 
\beqn\label{tuika-june-D}
\lefteqn{\cC_{\mbox{{\scriptsize\rm CSI-ED}}} \supset \bigcup_{p_Sp_{V}}}\non\\
& & \{{\mbox{rate pairs $(R_M, R_K)$ satisfying
(\ref{eq:keio-12}) and\ } (\ref{eq:keio-13-2})}\},\non\\
& & 
\eeqn

where 
ED denotes that the causal CSI $S$ is available at both Alice and Bob.
}
 \ekei
 
  {\em Proof:}  It is sufficient to replace $Y$ by $SY$ in (\ref{eq:keio-5}) and (\ref{eq:keio-6}). \IEEEQED.
 \bchui\label{chui:ED2}
 {\rm
 As far as we are concerned with ``degraded" WTCs ($Z$ is a degraded version of $Y$), the inclusion 
 $\supset$ in (\ref{tuika-june-D}) 
 can be replaced by $=$, so that in this case
  (\ref{tuika-june-D}) actually gives the causal SM-SK capacity region,
  as will be explicitly stated  later in Theorem \ref{teiri:4}.
  \IEEEQED
 }
 \echui
  \bchui\label{chui:sune-1}
{\rm 
Setting $R_M=0$ in  (\ref{tuika-june-D})
 yields one more new lower bound:
\beq\label{nx-2}
C^{{\scriptsize\sf K}}_{\mbox{{\scriptsize\rm CSI-ED}}} \ge\max_{p_Sp_V}( I(V;Y|S)-I(V;Z|S) +H(S|Z)).
\eeq
where  $V$ and $S$ are independent, and 
$C^{{\scriptsize\sf K}}_{\mbox{{\scriptsize\rm CSI-ED}}}$ denotes  the causal SK capacity.

On the other hand,
setting $R_K =0$  in (\ref{tuika-june-D}) yields the  lower bound given by 
Chia and El Gamal \cite[Theorem 1]{chia-elgamal}:
\beqn\label{nx-2-s}
C^{{\scriptsize\sf M}}_{\mbox{{\scriptsize\rm CSI-ED}}} &\ge &
\max_{p_Sp_V} \min ( I(V;Y|S)-I(V;Z|S) \non\\
& & \qquad\qquad\quad+H(S|Z),I(V;Y|S)),
\eeqn
with independent $V$ and $S$, where
$C^{{\scriptsize\sf M}}_{\mbox{{\scriptsize\rm CSI-ED}}}$ denotes  the causal SM capacity.
%
%
%
\IEEEQED
%
%
}
\echui
%
%
   %
 \bkei[Causal inner bound (3) at Alice and Bob]\label{kei:chia-el2}
 {\rm 
 Let us consider the WTC with causal CSI  at both Alice and Bob, as depicted in Fig. \ref{fig2}.
 Then, a pair $(R_M, R_K)$ is achievable if the following rate constraints are satisfied:
 \beqn\label{eq:non-p19}
R_M &\le& I(U;Y|S)\label{eq:chia-31}\\
R_M +R_K &\le& H(S|UZ),\label{eq:chia-41}
\eeqn
where $U$ and $S$ are independent, 
%
Thus,
 \beq\label{tuika-june-E}
\cC_{\mbox{{\scriptsize\rm CSI-ED}}} \supset \bigcup_{p_Sp_{U}}\{\mbox{rate pairs $(R_M, R_K)$ satisfying
(\ref{eq:chia-31}) and\ (\ref{eq:chia-41}})\}.
\eeq
%
 }
 \ekei

  {\em Proof:}  It is sufficient to replace $Y$ by $SY$ in (\ref{eq:keio-10}) and (\ref{eq:keio-10-2}). \IEEEQED

  \bchui\label{chui:sune-1}
{\rm 
Setting $R_K =0$  in (\ref{tuika-june-E})  
 yields the lower bound given by 
Chia and El Gamal \cite[Theorem 3]{chia-elgamal}: 
\beq\label{eq:asa-1}
C^{{\scriptsize\sf M}}_{\mbox{{\scriptsize\rm CSI-ED}}} \ge \max_{p_Sp_U}
\min (H(S|UZ), I(U;Y|S)).
\eeq
On the other hand, setting $R_M=0$ in  (\ref{tuika-june-E}) 
yields $C^{{\scriptsize\sf K}}_{\mbox{{\scriptsize\rm CSI-ED}}} \ge H(S|UZ)$. 
Also, we can set $U=\emptyset$ to 
obtain
\beq\label{eq:ma-1}
C^{{\scriptsize\sf K}}_{\mbox{{\scriptsize\rm CSI-ED}}} \ge \max_{p_{SX}}H(S|Z),
\eeq
which is obviously attained without transmission coding at the encoder, because in this case sharing of common secret key
at Alice and Bob is
enough without extra transmission of secret message (cf. Ahlswede and Csisz\'ar \cite{abc}).
Here, in view of (\ref{eq:ma-1}) and  \cite[Corollary 1]{khisti-wornell}, 
 it is easy to see that, for reversely degraded ($Y$ is a degraded version of $Z$) WTCs, 
\beq\label{eq:linden-1}
C^{{\scriptsize\sf K}}_{\mbox{{\scriptsize\rm CSI-ED}}} 
=C^{{\scriptsize\sf K}}_{\mbox{{\scriptsize\rm NCSI-ED}}}= \max_{p_{SX}}H(S|Z).
\eeq
%
}
\echui
   \bchui\label{chui:comp-1}
   {\rm
   Comparing (\ref{nx-2}) and (\ref{eq:ma-1}), we see that either one does not necessarily  subsume the other, 
   which depends on whether 
   $I(V;Y|S) \ge I(V;Z|S)$ or not.  Specifically, 
   in the case of $I(V;Y|S) \ge I(V;Z|S)$ coding helps, otherwise coding does not help. Notice that,
  for example, if $Z$ is a degraded version of $Y$,  then 
  $I(V;Y|S) \ge I(V;Z|S)$ always holds and so coding helps.
   \IEEEQED
   }
   \echui
\smallskip

{\em Case 3)} : Since $U=S\tilde{U}, V=\tilde{V}$ and $\tilde{U}\tilde{V}$ is independent of $S$, 
(\ref{eq:non-p2}) and (\ref{eq:non-p33}) are computed as 
\beqn
R_M &\le& I(\tU S\tV; Y)-I(\tU S\tV; S)\non\\
&=& I(\tU S\tV; Y) -H(S);\label{eq:le-1}
 %
 \eeqn
 \beqn
 R_M+R_K &\le& I(\tV;Y|S \tU)-I(\tV;Z|S\tU)\non\\
 & & -[I(S\tU;S) - I(S\tU;Y)]^+\non\\
 &=& I(\tV;Y|S\tU)-I(\tV;Z|S\tU)\non\\
 & & -[H(S)-I(S\tU;Y)]^+. \label{eq:le-2}
%
\eeqn
 As a consequence, removing tilde $\tilde{}$, we have the rate constraints,  with independent $UV$ and $S$,
\beqn
R_M &\le & I(USV;Y) -H(S);\label{eq:le-3}\\
R_M+R_K &\le &  I(V;Y|SU)-I(V;Z|SU)\non\\
 & & -[H(S)-I(SU;Y)]^+\label{eq:kan-1}.
\eeqn                                       
Therefore, any nonnegative rate pair $(R_M, R_K)$ is achievable if rate constraints 
(\ref{eq:le-3}) and (\ref{eq:kan-1}) are satisfied. Thus, we have the following one more fundamental inner bound (type II),
which is paired with Proposition \ref{prop:1} (type I):
\bmeidai[Causal SM-SK inner bound: type II]\label{prop:2}
 \beqn\label{tuika-june-R}
\lefteqn{\cC_{\mbox{{\scriptsize\rm CSI-E}}}} \non\\
&\supset & \bigcup_{p_Sp_{UV}}
\{{\mbox{rate pairs $(R_M, R_K)$ satisfying
(\ref{eq:le-3}) and\ } (\ref{eq:kan-1})}\}.\non\\
& & 
\eeqn
\emeidai
\bchui\label{chui:atto-1}
{\rm
We observe here that (\ref{eq:le-3}) and (\ref{eq:kan-1}) remain invariant under replacement of $Z$ by $SZ$.
This  implies that  the achievability due to Case 3) is invulnerable to the leakage of state information $S^n$ to Eve,
which is in notable  contrast with Case 2). 
\IEEEQED
}
\echui

An immediate consequence of (\ref{tuika-june-R})  is the following corollary:

\bkei[Causal lower bound (4) at Alice]\label{kei-poiu-3}
{\rm 
\beqn
C^{{\scriptsize\sf M}}_{\mbox{{\scriptsize\rm CSI-E}}} 
&\ge &\max_{p_Sp_{UV}} \min (I(V;Y|SU)-I(V;Z|SU)\non\\
& & -[H(S)-I(SU;Y)]^+,  I(USV;Y)-H(S)),\label{eq:silk-4}\non\\
\label{eq:silk-4}\\
C^{{\scriptsize\sf K}}_{\mbox{{\scriptsize\rm CSI-E}}} 
&\ge &\max_{\stackrel{p_Sp_{UV}:}{I(US V;Y)\ge H(S)}}
  (I(V;Y|SU)-I(V;Z|SU) \non\\
  & &\qquad\qquad\qquad-[H(S)-I(SU;Y)]^+),\label{eq:silk-5}
\eeqn
where $UV$ and $S$ are independent.
\IEEEQED
}
\ekei

{\em Proof:}  Setting $R_K=0$ in  (\ref{tuika-june-R})  yields (\ref{eq:silk-4}), while
setting $R_M=0$  in  (\ref{tuika-june-R})  yields (\ref{eq:silk-5}). \IEEEQED
%
%
%
\medskip

\bchui[Comparison of Case 2) and Case 3)]\label{chui:yet:1}
\mbox{}

{\rm
We  first notice  that (\ref{eq:le-3}) is the same as (\ref{eq:keio-3}), and moreover, noting that 
%
\beqn
H(S)-I(SU;Y)\non
&=& H(S|Y)-I(U;Y|S)\non\\
&=& H(S|Y) -I(U;SY)\non\\
&=& H(S|Y)-I(U;Y)-I(U;S|Y)\non\\
&=& H(S|UY)-I(U;Y)
\label{eq:keio-d1}
\eeqn 
and summarizing (\ref{eq:le-3}),  (\ref{eq:kan-1}) and (\ref{eq:keio-d1}), we have for Case 3).
\beqn
R_M &\le & I(USV;Y) -H(S);\label{eq:le-3a}\\
R_M+R_K &\le &  I(V;Y|SU)-I(V;Z|SU)\non\\
 & & -[H(S|UY)-I(U;Y)]^+.\label{eq:kan-1a}
\eeqn
In order to compare  this with that for Case 2), we rewrite (\ref{eq:keio-3}) 
 and (\ref{eq:keio-4-1})  as 
\beqn
R_M&\le& I(USV;Y) -H(S);\label{eq:le-4a}\\
R_M+R_K &\le& I(SV;Y|U)-I(SV;Z|U)\non\\
&=& I(S;Y|U)-I(S;Z|U)\non\\
& & +I(V;Y|SU)-I(V;Z|SU)\non\\
&=& I(V;Y|SU)-I(V;Z|SU) \non\\
& & -[H(S|UY)-H(S|UZ)].\non
\eeqn
Thus,  for Case 2),
\beqn
R_M &\le & I(USV;Y) -H(S);\label{eq:tuika-0}\\
R_M+R_K &\le &  I(V;Y|SU)-I(V;Z|SU)\non\\
 & & -[H(S|UY)-H(S|UZ)].\label{eq:tuika-1}
\eeqn

Comparing (\ref{eq:kan-1a}) and (\ref{eq:tuika-1}), we see that 
the difference consists in that of the terms $[H(S|UY)-I(U;Y)]^+$ and $[H(S|UY)-H(S|UZ)]$, so 
either one does not necessarily subsume the other, which depends on the choice of 
achievable probability distributions $p_{YZSXUV}.$
}
\IEEEQED
\echui

\begin{figure}[htbp]
\begin{center}
\includegraphics[width=80mm]{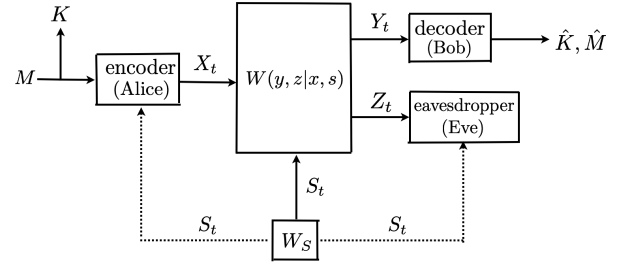}
\end{center}
\caption{WTC with the same CSI available at Alice and Eve ($t=1,2,\cdots, n$).
          }
\label{fig3}
\end{figure}

%
\bchui\label{chui:keio-111}
{\rm
As such, to get more insight, let us consider the WTC
with causal CSI available at both Alice and Eve, as 
depicted in Fig. \ref{fig3}. Then, since 
$[H(S|UY)-I(U;Y)]^+\le H(S|UY)$ and $[H(S|UY)-H(S|UZ)]=H(S|UY)$,  in this case Case 3)
outperforms Case 2), where $Z$ was replaced by $SZ$ as the state $S$ is available also at Eve
(cf. Remark \ref{chui:atto-1}).
This means that
Case 3) is preferable to Case 2) 
when Eve  have full access to  $S^n$.

On the other hand, consider an opposite case with CSI available at both Alice and Bob as n Fig. \ref{fig2}.
Then,  since $H(S|UY)=0$ with $SY$ instead of $Y$ and hence  
$[H(S|UY)-I(U;Y)]^+=0$ and $[H(S|UY)-H(S|UZ)]=-H(S|UZ)$, we see that,  in this case, 
 Case 2) outperforms Case 3).
}
\IEEEQED
\echui
\bchui\label{chui:keio-112}
{\rm
As is seen from the proof of Theorem \ref{teiri:1} in Bunin {\em et al.} \cite{bunin, bunin-2019}, 
in both cases of Case 2) and Case 3)
the state information $S^n$ is  to be reliably reproduced at Bob,
while the crucial difference between Case 2) and Case 3) is that in Case 2) 
 the $S^n$ is used  to carry  on secure transmission of message and/or key 
between Alice and Bob, whereas in Case 3) the $S^n$ is not used to convey  secure message and/or key
but  simply to help reliable (secured or unsecured)  transmission.  
On the other hand, in Case 1) the $S^n$ is not  to be reproduced at Bob.
As was illustrated in Remark \ref{chui:keio-111},
 favorable choices of these three cases depend on the probabilistic structure of WTCs.
 \IEEEQED
}
\echui


\ 
\smallskip
{\em Case 4)} : Since $U=S\tilde{U}, V=S\tilde{V}$ and $\tilde{U}\tilde{V}$ is independent of $S$, 
(\ref{eq:non-p2}) and (\ref{eq:non-p33}) are computed as 
\beqn
R_M &\le& I(\tU S\tV; Y)-I(\tU S\tV; S)\non\\
&=& I(\tU S\tV; Y) -H(S);\label{eq:le-1a4}\\
 %
 R_M+R_K &\le& I(S\tV;Y|S \tU)-I(S\tV;Z|S\tU)\non\\
 & & -[I(S\tU;S) - I(S\tU;Y)]^+\non\\
 &=& I(\tV;Y|S\tU)-I(\tV;Z|S\tU)\non\\
 & & -[H(S)-I(S\tU;Y)]^+, \label{eq:le-2a5}
%
\eeqn
which is nothing but (\ref{eq:le-1}) and (\ref{eq:le-2}) in Case 3), and therefore Case 4) reduces to Case 3).

\medskip

V.B: {\em Causal outer  bound:}

So far we have discussed a diversity of causal SM-SK inner bounds, but not about outer bounds.
This is because,  in general,  it is much harder with the problem of 
 {\em causal} outer bounds, in contrast with {\em non-causal} outer bounds.
 However, we can show an example of causal ``tighter"  outer bound, which is a rare case 
 (from the causal viewpoint) and is
 paired with Proposition \ref{prop:1} (achievability part).
In passing this section we consider this problem.

To do so, we first notice that the coding scheme used to prove Proposition \ref{prop:1} 
required the CSI $S^n$ to be reliably reproduced at Bob, i.e.,
$H(S^n|Y^n)\le n\vep_n$.
This kind of coding scheme is said to be {\em state-reproducing} (cf. Han and Sasaki \cite{han-sasaki-c}).
Then, one may ask what happens if we confine ourselves to within such state-reproducing coding schemes.
An answer is:
\bmeidai[Causal/non-causal outer bound]\label{teiri:yatta1}
{\rm
With state-reproducing coding schemes, we have the following  outer bound:
 \beqn\label{tuika-june-O-1}
\lefteqn{\cC_{\mbox{{\scriptsize\rm CSI-E}}}
 \subset 
\cC_{\mbox{{\scriptsize\rm NCSI-E}}}} \non\\
&\subset & \bigcup_{p_{SUV}}
\{{\mbox{rate pairs $(R_M, R_K)$ satisfying
(\ref{eq:non-p10}) and\ } (\ref{eq:keio-4-1})}\}\non\\
& &
\eeqn
}
\emeidai
Notice that the difference between  Proposition \ref{teiri:yatta1} (outer bound) and Proposition \ref{prop:1} (inner bound)  is that
the union in the former is taken over all probability distributions $p_{SUV}$'s, while 
in the latter the union is taken over all product probability distributions  $p_Sp_{UV}$'s. \IEEEQED

{\em Proof:}  
It suffices only to literally parallel the converse part  of Theorem \ref{teiri:3} with $\overline{Y}^n=S^nY^n$ replaced by $Y^n$, while
using 
$H(S^n|Y^n) \le n\vep_n$ (due to the state-reproducibility)  in inequality (\ref{eq:rev-1}) of Section \ref{sec:non-alice-bob},
which together with (\ref{eq:stte1}) (with $Y_t$ instead of $\overline{Y}_t$) brings about the required outer bound.
\IEEEQED
%
%

%

An immediate consequence  of (\ref{tuika-june-O-1})  is the following 
corollary, which is paired with Corollary \ref{kei-poiu-1}:

\bkei[Causal/non-causal upper bound]\label{kei-poiu-12}
{\rm 
With state-reproducing coding schemes, we have the upper bounds:
\beqn
C^{{\scriptsize\sf M}}_{\mbox{{\scriptsize\rm CSI-E}}} &\le &
C^{{\scriptsize\sf M}}_{\mbox{{\scriptsize\rm NCSI-E}}} \non\\
&\le & \max_{p_{SUV}}\min ( I(SV;Y|U)-I(SV;Z|U), \non\\
& &\qquad\qquad\qquad\quad I(U SV;Y)-H(S)),\label{eq:silk-12}\\
C^{{\scriptsize\sf K}}_{\mbox{{\scriptsize\rm CSI-E}}} &\le &
C^{{\scriptsize\sf K}}_{\mbox{{\scriptsize\rm NCSI-E}}} \non\\
&\le & \max_{\stackrel{p_{USV}:}{I(US V;Y)\ge H(S)}}
 (I(SV;Y|U)-I(SV;Z|U)).\non\\
 & &\label{eq:silk-22}
\eeqn
}
\ekei

\section{SM-SK Capacity Theorems for Degraded WTCs}
\label{sec-sececy-region}

\smallskip
%

\smallskip
%

 1) Let us now address the problem of SM-SK capacity regions 
to provide the exact  SM-SK capacity region for degraded WTCs with {\em causal/non-causal}  CSI
available  at ``both"  Alice and Bob as in Fig. \ref{fig2}. To do so, 
let the corresponding causal SM-SK capacity region 
be denoted by $\cC^{{\sf d}}_{\mbox{{\scriptsize\rm CSI-ED}}}$.
Similarly, the corresponding non-causal SM-SK capacity region 
is denoted by $\cC^{{\sf d}}_{\mbox{{\scriptsize\rm NCSI-ED}}}$.
%
Moreover, let  $\overline{\cR}^{{\sf d}}_{\sf in}(p_{SX})$ denote the set of 
all nonnegative rate pairs $(R_M, R_K)$ satisfying
the rate constraints:
\beqn
R_M &\le& I(X; Y|S),\label{eq:non2d-110}\\
R_M+R_K &\le& I(X;Y|S)-I(X;Z|S)+H(S|Z) \non\\
& &.\label{eq:non3d-2}
%
\eeqn
%
 Then, we have 
\bteiri[Causal/non-causal  SM-SK capacity region]\label{teiri:4}
Consider a degraded WTC ($Z$ is a degraded version of $Y$)
with causal/non-causal CSI  at Alice and Bob. Then,
\beqn\label{eq:non-4d-1}
\cC^{{\sf d}}_{\mbox{{\scriptsize\rm CSI-ED}}} &=&
\cC^{{\sf d}}_{\mbox{{\scriptsize\rm NCSI-ED}}}  \non\\
& =&\overline{\cR}^{{\sf d}}_{\sf in}
 \stackrel{\Delta}{=} 
 \bigcup_{p_{SX}} \overline{\cR}^{{\sf d}}_{\sf in}(p_{SX}),
\eeqn
where  the union  is taken over all  possible probability distributions $p_{SX}$'s.
\IEEEQED
\eteiri
%
%
\bchui\label{chui:hop-1}
{\rm
Notice, in particular,  that Theorem \ref{teiri:4} means also that the causal and non-causal capacity regions 
 coincide for degraded WTCs. 
 An immediate consequence of Theorem \ref{teiri:4} is 
}
 \IEEEQED
 \echui
 %
\bkei\label{kei:tuika-1}
{
\beqn
C^{{\sf d,M}}_{\mbox{{\scriptsize\rm CSI-ED}}} &=&
C^{{\sf d,M}}_{\mbox{{\scriptsize\rm NCSI-ED}}}  \non\\
&=& \max_{p_{SX}}\min(I(X;Y|S)-I(X;Z|S) \non\\
& & \qquad \qquad+H(S|Z), I(X;Y|S)),
\label{eq:hop-2}\\
C^{{\sf d,K}}_{\mbox{{\scriptsize\rm CSI-ED}}} &=&
C^{{\sf d,K}}_{\mbox{{\scriptsize\rm NCSI-ED}}}  \non\\
&=&\max_{p_{SX}}
(I(X;Y|S)-I(X;Z|S) \non\\
& &\qquad\qquad+H(S|Z)),\label{eq:hop-3}
\eeqn
where 
$C^{{\sf d,M}}_{\mbox{{\scriptsize\rm CSI-ED}}}, 
C^{{\sf d,M}}_{\mbox{{\scriptsize\rm NCSI-ED}}}$ 
$\left({\rm resp}. \  C^{{\sf d,K}}_{\mbox{{\scriptsize\rm CSI-ED}}}, 
C^{{\sf d,K}}_{\mbox{{\scriptsize\rm NCSI-ED}}} \right)$
is the supremum of  the projection of $\cC^{{\sf d}}_{\mbox{{\scriptsize\rm CSI-ED}}},
\cC^{{\sf d}}_{\mbox{{\scriptsize\rm NCSI-ED}}}$
on the $R_M$-axis (resp. $R_K$-axis).\label{eq:hop-4}
\IEEEQED
}
\ekei

\bchui
Formula (\ref{eq:hop-2})  has earlier been given by  \cite[Theorem 3]{chia-elgamal}
in a quite different manner.
\IEEEQED
\echui
%

\medskip

{\em Proof of achievability for Theorem \ref{teiri:4}:}

Let $(X,S)$ be arbitrarily given, then the functional representation lemma \cite{gamal-kim} claims that
there exist a random variable $V$ and a deterministic function $f: \cV\times \cS\to \cX$
such that $V$ and $S$ are independent and $X=f(V, S)$. Then, Theorem \ref{teiri:2} (Case 2) $A$: 
with $U=\emptyset$) claims
that any rate pair $(R_M, R_K)$ satisfying the rate constraints (\ref{eq:keio-12}) and (\ref{eq:keio-13-2}), that is, 
\beqn
 R_M &\le & I(V;Y|S);\label{eq:keio-5d}\\
R_M+R_K &\le&  I(V;Y|S)-I(V;Z|S) \non\\
& & \qquad\qquad\qquad+H(S|Z),\label{eq:keio-6d}
\eeqn
is ``causally" achievable. Then, it suffices to observe that 
the right-hand sides of (\ref{eq:keio-5d})  
is rewritten as 
\beqn
\lefteqn{ I(V;Y|S)}\non\\
&\stackrel{(e)}{=}& I(VX;Y|S)\non\\
 &\stackrel{(g)}{=}& I(X;Y|S),\label{eq:haya-1}
 \eeqn
 where $(e)$ is because $X$ is a deterministic function of $(V, S)$;
$(g)$ follows from the Markov chain property $UV\to SX \to YZ$.
 Similarly,  (\ref{eq:keio-6d})) can be rewritten as 
 \beqn
\lefteqn{ I(V;Y|S)-I(V;Z|S) +H(S|Z)}\non\\
  &=& I(X;Y|S) -I(X;Z|S) 
  +H(S|Z).\label{eq:haya-2}
%
\eeqn

\noindent
{\em Proof of converse for Theorem \ref{teiri:4}:}

Theorem \ref{teiri:3} claims that any achievable rate pair $(R_M, R_K)$ must satisfy
the rate constraints (\ref{eq:non2d}) and (\ref{eq:non3d}), that is,
\beqn
R_M &\le& I(UV; Y|S),\label{eq:non2d-1}\\
R_M+R_K &\le& I(V;Y|SU)-I(V;Z|SU) \non\\
& &\qquad\qquad +H(S|ZU)\label{eq:non3d-2z}
%
\eeqn
with some $UVSXYZ$.
%
The right-hand sides of (\ref{eq:non2d-1}) and (\ref{eq:non3d-2z})  are evaluated as  follows:
\beqn
I(UV;Y|S) 
&\le& I(UVX;Y|S)\non\\
 &=&I(X;Y|S) +I(UV;Y|SX)\non\\
  &\stackrel{(v)}{=}&I(X;Y|S),\label{eq:mata-q}
  \eeqn
  where $(v)$ follows from the Markov chain property $UV\to SX\to Y$.
  Hence,
  \beq\label{eq:bon-1}
  I(UV;Y|S) \le I(X;Y|S).
  \eeq
  %
 %
 On the other hand, 
 \beqn
  \lefteqn{I(V;Y|SU)  - I(V;Z|SU)}\non\\ 
  &=& I(VX;Y|SU)-I(X;Y|SUV)\non\\
  && - I(VX;Z|SU)+I(X;Z|SUV)\non\\
&=& I(VX;Y|SU)-I(VX;Z|SU) \non\\
  & &  - [I(X;Y|SUV)-I(X;Z|SUV)]\non\\
   &\stackrel{(a)}{=}& I(X;Y|SU)-I(X;Z|SU) \non\\
  & &  - [I(X;Y|SUV)-I(X;Z|SUV)]\non\\
    &\stackrel{(b)}{\le} & I(X;Y|SU)-I(X;Z|SU) \non\\
&=&I(UX;Y|S)-I(UX;Z|S)\non\\
     & &-[I(U;Y|S)-I(U;Z|S)]\non\\
     &\stackrel{(c)}{\le} &I(X;Y|S)-I(X;Z|S)\non\\
     & &-[I(U;Y|S)-I(U;Z|S)],
        \label{eq:utuo-1}
 \eeqn
where
$(a), (c)$ follows from the Markov chain property $UV\to SX\to YZ$; 
$(b)$ follows from the assumed degradedness. 
Moreover, since
\beq\label{preq:1}
H(S|ZU) -H(S|Z) = - I(S;U|Z),
\eeq
it follows that
\beqn
\lefteqn{H(S|ZU) -H(S|Z)
-[I(U;Y|S)-I(U;Z|S)]}
\non\\
&=& - I(S;U|Z) 
-[I(U;Y|S)-I(U;Z|S)]\non\\
&=& - I(U;Y|S) - [I(S;U|Z) -I(U;Z|S)]\non\\
&=& I(S;U)-I(U;SY) -[ I(S;U)-I(U;Z)]\non\\
&=& -I(U;SY)+I(U;Z)\non\\
&\le  & -I(U;Y)+I(U;Z)\non\\
&\stackrel{(j)}{=}&- I(U;Y|Z) \le 0,
\eeqn
where $(j)$ follows from the assumed degradedness.
Therefore, 
\beqn\label{peq:den1}
\lefteqn{H(S|ZU)}\non\\
& & -[I(U;Y|S)-I(U;Z|S)]\non\\
& &\le  H(S|Z)).
\eeqn
Thus, by virtue of  
(\ref{eq:utuo-1}) and (\ref{peq:den1}), we obtain
\beqn\label{eq:natto-2}
  \lefteqn{I(V;Y|SU)-I(V;Z|SU)+ H(S|ZU)}\non\\ 
  &\le & I(X;Y|S)-I(X;Z|S) +H(S|Z),
  \eeqn
  which together with  (\ref{eq:bon-1}) 
completes the proof of Theorem \ref{teiri:4}.
%

\smallskip

2) Next let us  address the problem of SM-SK capacity region 
to provide an   SM-SK outer bound for degraded WTCs with {\em causal/non-causal}  CSI
available ``only" at  Alice as in Fig.\ref{fig1}. To do so, 
let the corresponding causal SM-SK capacity region 
be denoted by $\cC^{{\sf e}}_{\mbox{{\scriptsize\rm CSI-E}}}$.
Similarly, the corresponding non-causal SM-SK capacity region 
is denoted by $\cC^{{\sf e}}_{\mbox{{\scriptsize\rm NCSI-E}}}$.
Moreover, let  $\overline{\cR}^{{\sf e}}_{\sf out}(p_{SX})$ denote the set of 
all nonnegative rate pairs $(R_M, R_K)$ satisfying
the rate constraints:
\beqn
R_M &\le& I(X; Y|S),\label{eq:non2d-11}\\
R_M+R_K &\le& I(X;Y|S)-I(X;Z|S) \non\\
& & \qquad+H(S|Z)-H(S|Y).\label{eq:non3d-21}
\eeqn
Then, we have

\bteiri[Causal/non-causal  SM-SK outer bound]\label{teiri:5}
Consider a degraded WTC ($Z$ is a degraded version of $Y$)
with causal/non-causal CSI  at Alice. Then,
\beqn\label{eq:non-4d-1}
\cC^{{\sf e}}_{\mbox{{\scriptsize\rm CSI-E}}} & \subset &
\cC^{{\sf e}}_{\mbox{{\scriptsize\rm NCSI-E}}}  \non\\
& \subset &\overline{\cR}^{{\sf e}}_{\sf out}
 \stackrel{\Delta}{=} 
 \bigcup_{p_{SX}} \overline{\cR}^{{\sf e}}_{\sf out}(p_{SX}),
\eeqn
where  the union  is taken over all  possible probability distributions $p_{SX}$'s.
\IEEEQED
\eteiri

{\em Proof:}  
The upper bound  
 (\ref{eq:non2d-110}) for  WTCs with CSI at both Alice and Bob
 must hold also for  WTCs with CSI only at Alice, yielding (\ref{eq:non2d-11}).

On the other hand, in order to yield inequality (\ref{eq:non3d-21}), 
it suffices to  parallel the converse proof of Theorem \ref{teiri:3} 
while keeping   
in mind $H(S^n|MK\overline{Y}^n)\le H(S^n|MKZ^n)$ with $Y^n$ instead of $\overline{Y}^n=S^nY^n$
(due to the assumed degradedness) 
 in (\ref{eq:sune10}) 
 and skipping  (\ref{eq:rev-1}) 
 to (\ref{eq:sune19})  $\sim$ (\ref{eq:br1}) (with $Y_t$ instead of $\overline{Y}_t$), 
which 
claims that
the achievable rate pair $(R_M, R_K)$ needs to satisfy
the rate constraints:
\beqn
\lefteqn{n(R_M+R_K)}\non\\
 &\le &\sum_{t=1}^n I(V_t;Y_t|S_tU_t)-\sum_{t=1}^nI(V_t;Z_t|S_tU_t) \non\\
& & +\sum_{t=1}^nH(S_t|Z_tU_t)-\sum_{t=1}^nH(S_t|Y_tU_t)+4n\vep_n\non\\
&=& nI(V;Y|SU)-nI(V;Z|SU)\non\\
& &  +nH(S|ZU)-nH(S|YU) +4n\vep.
 \label{eq:non3d-2}
%
\eeqn
Therefore, by dividing by $n$ and letting $n\to\infty$, we have 
\beqn\label{eq:iti-eno8}
R_M+R_K & \le& 
I(V;Y|SU)-I(V;Z|SU)\non\\
& &  +H(S|ZU)-H(S|YU).
\eeqn
Then, in the same manner as in the converse proof of Theorem \ref{teiri:4}, 
we can check that (\ref{eq:iti-eno8}) yields inequality (\ref{eq:non3d-21}).
\IEEEQED

\medskip
Finally, the following corollary  follows from Theorem \ref{teiri:5}:

\bkei[Upper bound on SK rates]\label{kei:tesun} For a degraded WTC
with causal/non-causal CSI  at Alice, we have
\beqn\label{eq:hantesun}
\lefteqn{C^{{\sf e,K}}_{\mbox{{\scriptsize\rm CSI-E}}}
\le C^{{\sf e,K}}_{\mbox{{\scriptsize\rm NCSI-E}}}}\non\\
 &\le& \max_{p_{SX}}(I(X;Y|S)-I(X;Z|S) +H(S|Z)-H(S|Y)).\non\\
 & &
\eeqn
\ekei

\section{Concluding Remarks}\label{sec-revisited-remark}

So far, we have studied the coding problem for WTCs with causal/non-causal CSI available at Alice 
and/or Bob under the semantic security criterion, the key part of which was summarized  as Theorem \ref{teiri:2}
for WTCs with {\em causal} CSI at Alice.
As is already clear, all the advantages of Theorem \ref{teiri:2} are inherited directly from Theorem \ref{teiri:1}
that had been established by Bunin {\em et al.} \cite{bunin-2019} for WTCs with {\em non-causal} CSI at Alice, 
This suggests that it is sometimes useful to deal with the causal problem as a special class of  non-causal problems.

It is rather surprising  to see that all the previous results \cite{chia-elgamal}, \cite{fujita}, \cite{han-sasaki-c} 
for WTCs with {\em causal} CSI follow immediately from Theorem  \ref{teiri:2} alone.
 Notice here that the validity of Theorem  \ref{teiri:1} 
 is based heavily 
 on the superiority of  the two layered superposition coding scheme (cf. \cite{prabhakaran}, \cite{goldfeld})
 along with that of soft covering lemma.
It is also pleasing  to see that Theorem \ref{teiri:3}, as a by-product of Theorem \ref{teiri:1},
gives for the first time  the exact SM-SK  capacity region for WTCs with non-causal CSI
at both Alice and Bob. Theorem \ref{teiri:4}  is also regarded as one of the key results from the viewpoint of SK-SM capacity regions
for degraded WTCs.

\begin{figure}[htbp]
\begin{center}
\includegraphics[width=80mm]{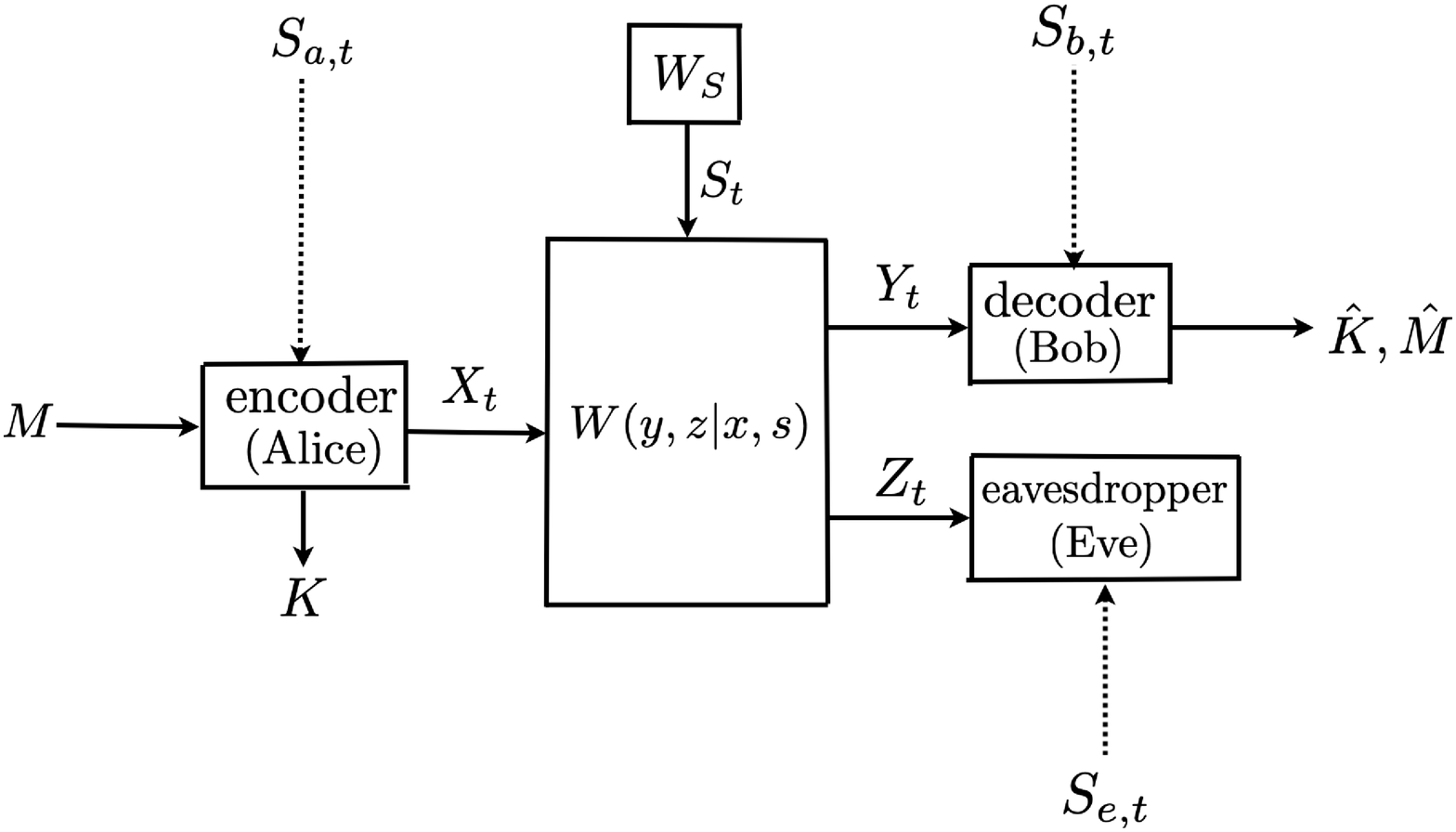}
\end{center}
\caption{WTC with  causal CSIs $S_a, S_b, S_e$ available at Alice, Bob  and Eve ($t=1,2,\cdots, n$).
          }
\label{fig4}
\end{figure}

Although Theorem \ref{teiri:2} treats the WTC with causal CSI available only at Alice, 
it can actually be effective also for investigating general WTCs with three correlated causal CSIs 
$S_a, S_b, S_e$ (correlated with state  $S$)
available at Alice, Bob and Eve, respectively (cf. Fig. \ref{fig4}).

 We would like to remind that this seemingly “general” WTCs  actually boils down 
  to the so far studied WTC with causal CSI available only at Alice 
  simply by replacing channel $W_{YZ|SX}(y,z|s,x)$ with $W_{YZ|S_aX}(y,z|s_a, x)$
 $\stackrel{\Delta}{=}\sum_{s}W_{YZ|SX}(y,z|s,x)p(s|s_a)$
 and at the same time 
   by replacing $Y, Z$ with $S_bY, S_eZ$, respectively,
  %
%
 In this connection, the reader may refer, for example,
  to Khisti, Diggavi and  Wornell \cite{khisti-wornell}, and Goldfeld, Cuff and Permuter \cite{goldfeld}.

\appendices

%
%
%
%

\section{Proof of Lemma \ref{hodai:naotta}}\label{addA}
From the manner of generating the random code, we see 
   that the total joint probability of all $(\ssu_i,\ssv_{ij})'$s is given by $P_{1n}P_{2n}P_{3n},$
where
\beqn
P_{1n}
&=&\prod_{k=2}^{L_n}\prod_{\ell =1}^{N_n} p(\ssu_k)p(\ssv_{k\ell} |\ssu_k),\label{eq:gsypr1}\\
P_{2n}&=&
\prod_{\ell =2}^{N_n} p(\ssv_{1\ell} |\ssu_1),\label{eq:gsypr13}\\
P_{3n} &=& p(\ssu_1, \ssv_{11}).\label{eq:elem20}
\eeqn
We now  directly develop $\Exp D(q^n_S||p^n_S)$ as follows. Here, for simplicity, we set $p(\sss)=p^n_S(\sss)$. 
\beqn
\lefteqn{\Exp D(q^n_S||p^n_S) }\non\\
&=& \sum_{\sss\in \cS^n}
\sum_{i=1}^{L_n}\sum_{j=1}^{N_n}\sum_{\ssu_i\in \cU^n} \sum_{\ssv_{ij}\in \cV^n}
P_{1n}P_{2n}P_{3n}\non\\
& &\cdot \left(\frac{1}{L_nN_n}\sum_{i^{\prime}=1}^{L_n}\sum_{j^{\prime}=1}^{N_n}W(\sss|\ssu_{i^{\prime}}, 
\ssv_{i^{\prime}j^{\prime}})\right)\non\\
& & \cdot\log \left(\frac{1}{L_nN_np(\sss)}\sum_{k^{\prime}=1}^{L_n}\sum_{\ell^{\prime}=1}^{N_n}
W(\sss|\ssu_{k^{\prime}}, \ssv_{k^{\prime}\ell^{\prime}})\right)\non\\
&\stackrel{(a)}{=}&
\sum_{\sss\in \cS^n}
\sum_{i=1}^{L_n}\sum_{j=1}^{N_n}\sum_{\ssu_i\in \cU^n} \sum_{\ssv_{ij}\in \cV^n}
P_{1n}P_{21n}P_{3n}\non\\
& & \cdot W(\sss|\ssu_{1}, 
\ssv_{11})
\log \left(\frac{1}{L_nN_np(\sss)}\sum_{k^{\prime}=1}^{L_n}\sum_{\ell^{\prime}=1}^{N_n}
W(\sss|\ssu_{k^{\prime}}, \ssv_{k^{\prime}\ell^{\prime}})\right),\non\\
& &
 \label{eq:syouki-11}
\eeqn
where $(a)$ follows from the symmetry of codes. We  decompose the quantities in (\ref{eq:syouki-11})
 as
\beq
\sum_{k^{\prime}=1}^{L_n}\sum_{\ell^{\prime}=1}^{N_n}
W(\sss|\ssu_{k^{\prime}}, \ssv_{k^{\prime}\ell^{\prime}})= A_{1n}+A_{2n}+A_{3n},\label{eq:elene2}
\eeq
where
\beqn
A_{1n}&=& \sum_{k^{\prime}=2}^{L_n}\sum_{\ell^{\prime}=1}^{N_n}
W(\sss|\ssu_{k^{\prime}}, \ssv_{k^{\prime}\ell^{\prime}})\label{eq:elem3}\\
 A_{2n}&= & \sum_{\ell^{\prime}=2}^{N_n}
W(\sss|\ssu_1, \ssv_{1\ell^{\prime}})\label{eq:elem4}\\
 A_{3n}&=& 
W(\sss|\ssu_1, \ssv_{11}).\label{eq:elem5}
\eeqn
%
%
Again, from  the manner of generating  the random code, we see that $A_{1n}$ and $(A_{2n}, A_{3n})$ 
are independent, whereas $A_{2n}$ and $A_{3n}$ are conditionally independent given $\ssu_1$. Thus,
\beqn
\lefteqn{\Exp D(q^n_S||p^n_S) }\non\\
&=&
\sum_{\sss\in \cS^n}
\sum_{i=1}^{L_n}\sum_{j=1}^{N_n}\sum_{\ssu_i\in \cU^n} \sum_{\ssv_{ij}\in \cV^n}
P_{1n}P_{2n}P_{3n} \non\\
& & \cdot W(\sss|\ssu_{1}, 
\ssv_{11})
\log \left(\frac{A_{1n}+A_{2n}+A_{3n}}{L_nN_np(\sss)}\right)\non\\
&\stackrel{(b)}{\le}&
\sum_{\sss\in \cS^n}
\sum_{i=1}^{1}\sum_{j=1}^{N_n}\sum_{\ssu_i\in \cU^n} \sum_{\ssv_{ij}\in \cV^n}
P_{2n}P_{3n}\non\\
& & \cdot W(\sss|\ssu_{1}, 
\ssv_{11})
\log \left(\frac{\sum^{*} A_{1n}+A_{2n}+A_{3n}}{L_nN_np(\sss)}\right),\non\\
& & 
\label{eq:syouki-1}
\eeqn
where $(b)$ follows from the concavity of the function $x\mapsto \log x$ along with the Jensen's inequality.
Here, 
\beqn\label{eq:shouga-5}
\sum^{*} A_{1n} &\stackrel{\Delta}{=}& \sum_{i=2}^{L_n}\sum_{j=1}^{N_n}
\sum_{\ssu_i\in \cU^n} \sum_{\ssv_{ij}\in \cV^n}
P_{1n}A_{1n}\non\\
&=& (L_n-1)N_n p(\sss).
\eeqn
Hence,
 \beqn
 \lefteqn{\Exp D(q^n_S||p^n_S) }\non\\
 &\le&
\sum_{\sss\in \cS^n}
\sum_{i=1}^{1}\sum_{j=1}^{N_n}\sum_{\ssu_i\in \cU^n} \sum_{\ssv_{ij}\in \cV^n}
P_{2n}P_{3n}\non\\
& & \cdot W(\sss|\ssu_{1}, 
\ssv_{11})
\log \left( 1+\frac{A_{2n}+A_{3n}}{L_nN_np(\sss)}\right).
\label{eq:syouki-5}
 \eeqn
 Moreover, 
 \beqn\label{eq:syouki-7}
 \lefteqn{\Exp D(q^n_S||p^n_S) }\non\\
 &\le&
\sum_{\sss\in \cS^n}
\sum_{i=1}^{1}\sum_{j=1}^{1}\sum_{\ssu_i\in \cU^n} \sum_{\ssv_{ij}\in \cV^n}
P_{3n}\non\\
& & \cdot W(\sss|\ssu_{1}, 
\ssv_{11})
\log \left( 1+\frac{\sum^*A_{2n}+A_{3n}}{L_nN_np(\sss)}\right),
\label{eq:syouki-5}
 \eeqn
 where 
\beqn\label{eq:syouki-8}
\sum^*A_{2n} 
&\stackrel{\Delta}{=}& \sum_{i=1}^{1}\sum_{j=2}^{N_n}
\sum_{\ssu_i\in \cU^n} \sum_{\ssv_{ij}\in \cV^n}
P_{2n}A_{2n}\non\\
&=& (N_n -1)W(\sss|\ssu_1),
\eeqn
so that, with $0\le \rho <1$,
\beqn\label{eq:syouki-9}
 \lefteqn{\Exp D(q^n_S||p^n_S) }\non\\
 &\le&
\sum_{\sss\in \cS^n}
\sum_{i=1}^{1}\sum_{j=1}^{1}\sum_{\ssu_i\in \cU^n} \sum_{\ssv_{ij}\in \cV^n}
P_{3n}\non\\
& & \cdot W(\sss|\ssu_{1}, 
\ssv_{11})\non\\
& & \cdot
\log \left( 1+\frac{W(\sss|\ssu_1)}{L_np(\sss)}+\frac{W(\sss|\ssu_1, \ssv_{11})}{L_nN_np(\sss)}\right)\non\\
%
&=&
\sum_{\sss\in \cS^n}
\sum_{\ssu_1\in \cU^n} \sum_{\ssv_{11}\in \cV^n}
 p(\ssu_1, \ssv_{11})W(\sss|\ssu_{1}, 
\ssv_{11})\non\\
& & \cdot\log \left( 1+\frac{W(\sss|\ssu_1)}{L_np(\sss)}+\frac{W(\sss|\ssu_1, \ssv_{11})}{L_nN_np(\sss)}\right)\non\\
&=&
\sum_{\sss\in \cS^n}
\sum_{\ssu_1\in \cU^n} \sum_{\ssv_{11}\in \cV^n}
p(\sss, \ssu_1, \ssv_{11})
\non\\
& & \cdot 
\log \left( 1+\frac{W(\sss|\ssu_1)}{L_np(\sss)}+\frac{W(\sss|\ssu_1, \ssv_{11})}{L_nN_np(\sss)}\right)\non\\
&=& 
\sum_{\sss\in \cS^n}
\sum_{\ssu_1\in \cU^n} \sum_{\ssv_{11}\in \cV^n}
\frac{1}{\rho}p(\sss, \ssu_1, \ssv_{11})
\non\\
& & \cdot
\log \left( 1+\frac{W(\sss|\ssu_1)}{L_np(\sss)}+\frac{W(\sss|\ssu_1, \ssv_{11})}{L_nN_np(\sss)}\right)^{\rho}\non\\
&\stackrel{(c)}{\le}& 
\sum_{\sss\in \cS^n}
\sum_{\ssu_1\in \cU^n} \sum_{\ssv_{11}\in \cV^n}
\frac{1}{\rho}p(\sss, \ssu_1, \ssv_{11})
\non\\ 
& &  \cdot
\log \left( 1+\left(\frac{W(\sss|\ssu_1)}{L_np(\sss)}\right)^{\rho}+
\left(\frac{W(\sss|\ssu_1, \ssv_{11})}{L_nN_np(\sss)}\right)^{\rho}\right)\non\\
&\stackrel{(d)}{\le}&
\sum_{\sss\in \cS^n}
\sum_{\ssu_1\in \cU^n} 
\frac{1}{\rho}p(\sss, \ssu_1)\left(\frac{W(\sss|\ssu_1)}{L_np(\sss)}\right)^{\rho} \label{eq:moui-1}\\
& & +\sum_{\sss\in \cS^n}
\sum_{\ssu_1\in \cU^n} \sum_{\ssv_{11}\in \cV^n}\frac{1}{\rho}p(\sss, \ssu_1, \ssv_{11})
\left(\frac{W(\sss|\ssu_1, \ssv_{11})}{L_nN_np(\sss)}\right)^{\rho}.
\non\\
 & &\label{eq:moui-2}
\eeqn
%
where $(c)$ follows from $(x+y+z)^{\rho}\le x^{\rho} +y^{\rho}+z^{\rho}$;
$(d)$ follows from $\log (1+x)\le x.$ %
 For simplicity, we delete the subscripts $``1, 11"$ in (\ref{eq:moui-1}) and  (\ref{eq:moui-2})  to obtain
 \beqn\label{eq:gyos-1}
F_{1n} &\stackrel{\Delta} {=}&
\sum_{\sss\in \cS^n}
\sum_{\ssu\in \cU^n} 
\frac{1}{\rho}p(\sss, \ssu)\left(\frac{W(\sss|\ssu)}{L_np(\sss)}\right)^{\rho}, \label{eq:moui-31}\\
F_{2n} &\stackrel{\Delta} {=}&
\sum_{\sss\in \cS^n}
\sum_{\ssu\in \cU^n} \sum_{\ssv \in \cV^n}\frac{1}{\rho}p(\sss, \ssu, \ssv)
\left(\frac{W(\sss|\ssu, \ssv)}{L_nN_np(\sss)}\right)^{\rho}.\non\\
& & \label{eq:moui-2g2}
\eeqn

Hereafter, let us show that  $F_{1n}\to 0$, $ F_{2n}\to 0$ as $n$ tends to $\infty$ 
if rate constraints $R_1>I((U;S), R_1+R_2> I(UV;S)$ are satisfied.
%
First, let us show $F_{2n}\to 0$.
Since $p(\sss, \ssu, \ssv)=p(\ssu, \ssv)W(\sss|\ssu, \ssv)$,    $F_{2n}$  can be rewritten as
\beqn
\lefteqn{F_{2n}}\non\\
&=& \frac{1}{\rho (L_nN_n)^{\rho}}\sum_{\sss\in \cS^n}
\sum_{\ssu\in \cU^n} \sum_{\ssv \in \cV^n}p(\ssu, \ssv)
W(\sss|\ssu, \ssv)^{1+\rho}p(\sss)^{-\rho}.\non\\
& & \label{eq:moui-2g3}
\eeqn
On the other hand, by virtue of H\"older's inequality,
\beqn\label{eq:simple}
\lefteqn{\left(\sum_{(\ssu, \ssv)\in \cU^n\times \cV^n}p(\ssu, \ssv)W(\sss|\ssu, \ssv)^{1+\rho}\right)p(\sss)^{-\rho}} \non \\
&=& \left(\sum_{(\ssu, \ssv)\in \cU^n\times \cV^n}p(\ssu, \ssv)W(\sss|\ssu, \ssv)^{1+\rho}\right)
\non\\
& & \cdot
 \left(\sum_{(\ssu, \ssv)\in \cU^n\times \cV^n}p(\ssu, \ssv)W(\sss|\ssu, \ssv)\right)^{-\rho}\non\\
&\le& \left(\sum_{(\ssu, \ssv)\in \cU^n\times \cV^n} p(\ssu, \ssv)W(\sss|\ssu, \ssv)^{\frac{1}{1-\rho}}\right)^{1-\rho}
\eeqn
for  $0 < \rho < 1$. Therefore, it follows from (\ref{eq:moui-2g3}) that
\beqn\label{eq:app-8}
\lefteqn{F_{2n}}\non\\
&\le & \frac{1}{\rho (L_nN_n)^{\rho}} \sum_{\sss\in \cS^n}
 \left(\sum_{(\ssu, \ssv)\in \cU^n\times \cV^n}p(\ssu, \ssv)W(\sss|\ssu, \ssv)^{\frac{1}{1-\rho}}\right)^{1-\rho} \non\\
 &=& \frac{1}{\rho}\exp\left[
 -[n\rho(R_1+R_2)+ E_0(\rho, p)]\right],
\eeqn
where
\beqn\label{eq:app-9}
\lefteqn{E_0(\rho, p)} \non\\
&=& -\log\left[
 \sum_{\sss\in \cS^n}
 \left(\sum_{(\ssu, \ssv)\in \cU^n\times \cV^n}p(\ssu, \ssv)W(\sss|\ssu, \ssv)^{\frac{1}{1-\rho}}\right)^{1-\rho}
 \right].
 \non\\
 & &
 \eeqn
Then, by means of Gallager \cite[Theorem 5.6.3]{gall},  we have $E_0(\rho, p)|_{\rho=0}=0$ and 
\beqn\label{eq:app-10}
\left.\frac{\partial E_0(\rho, p)}{\partial \rho}\right|_{\rho=0}
&=&
-I(p, W)\non\\
&=&-I(\sU\sV; \sS)\non\\
&\stackrel{(e)}{=}& -nI(UV;S), \label{eq:ker-q}
\eeqn
where $(e)$ follows because $(\sU\sV;  \sS)$ is a correlated i.i.d. sequence with generic variable $(UV, S)$.
Thus, for any small constant $\tau>0$ there exists a $\rho_0>0$ such that,
for all $0<\rho \le \rho_0$,
\beq\label{eq:app-11}
E_0(\rho, p)\ge -n\rho(1+\tau)I(UV;S)
\eeq
which is substituted into (\ref{eq:app-8}) to obtain
\beqn\label{eq:app-12}
F_{2n}
&\le &  \frac{1}{\rho}\exp\left[
 -n\rho(R_1+R_2 -(1+\tau) I(UV;S))\right].\non\\
 & &
\eeqn
On the other hand, in view of  rate constraint $R_1+R_2>I(UV; S)$,  with some $\delta>0$ we can write 
\beq\label{eq:app-14}
R_1+R_2=I(UV;S)+2\delta,
\eeq
which leads to 
\beqn\label{eq:app-13}
\lefteqn{R_1+R_2 -(1+\tau)I(UV;S)}\non\\
&=& I(UV;S) +2\delta
 - I(UV;S) -
  I(UV;S)\non\\
&=& 2\delta -\tau I(UV;S).
\eeqn
We notice here that $\tau>0$ can be arbitrarily small,
so that the last term on the right-hand side of (\ref{eq:app-13})  can be made larger than $\delta>0$.
Then, (\ref{eq:app-12}) yields
\beq\label{eq:app-15}
F_{2n}\le \frac{1}{\rho}\exp[-n\rho\delta],
\eeq
which implies that with any small $\vep>0$ it holds that
\beq\label{eq:app-15}
F_{2n} \le \vep
\eeq
for all sufficiently large $n$. 

Similarily,
$F_{1n} \le \vep$ with rate  constraint $R_1>I(U; S)$ can also be shown.

Thus, 
 the proof of Lemma \ref{hodai:naotta} has been completed.
 \IEEEQED 

\section{Proof of Remark \ref{chui:sune3}}\label{addB}

%
%

For simplicity, set the right-hand sides of (\ref{eq:silk-10}), \ref{eq:silk-20}), (\ref{hasa-2}) and (\ref{eq:riri7}) as 
\beqn
M_1 &=& \max_{p_Sp_V}\min ( I(SV;Y)-I(SV;Z),\non\\
& &\qquad\qquad\qquad I(SV;Y)-H(S)),\label{eq:silk-10q}\\
K_1 &=&  \max_{\stackrel{p_Sp_{V}:}{I(S V;Y)\ge H(S)}}
 (I(SV;Y)-I(SV;Z)),\label{eq:silk-20q}\\
M_2 &=&  \max_{p_Sp_U}
\min(H(S|UZ)-H(S|UY), \non\\
& &\qquad\qquad\qquad I(U;Y)-H(S|UY)), \label{hasa-2q}\\
K_2 &=& \max_{\stackrel{p_Sp_U:}{I(U;Y)\ge H(S|UY)}}(H(S|UZ)-H(S|UY)),
\non\\ &&\label{eq:riri7q}
\eeqn
which, after some calculation  with $Y$ replaced by $SY$, leads, respectively,  to
\beqn
M_1^{\prime} &=& \max_{p_Sp_V}\min (I(V;Y|S)-I(V;Z|S) +H(S|Z), \non\\
& &\qquad\qquad\qquad I(V;Y|S)),\label{eq:silk-10q1}\\
K_1^{\prime} &=&  \max_{p_Sp_V}
 (I(V;Y|S)-I(V;Z|S) +H(S|Z)), \label{eq:silk-20q1}\\
M_2^{\prime} &=&  \max_{p_Sp_U}
\min(H(S|UZ),  I(U;Y|S)), \label{hasa-2q1}\\
K_2^{\prime} &=& \max_{p_Sp_U}H(S|UZ)=\max_{p_{SX}}H(S|Z).\label{eq:riri7q1}
\eeqn
Notice here that (\ref{eq:silk-10q})  $\sim$ (\ref{eq:riri7q}) give lower bounds for $\cC_{\mbox{{\scriptsize\rm CSI-E}}}$
(with CSI $S$ available only at  Alice),
 whereas (\ref{eq:silk-10q1}) $\sim$ (\ref{eq:riri7q1}) give lower bounds for $\cC_{\mbox{{\scriptsize\rm CSI-ED}}}$
 (with CSI $S$ available at both Alice and Bob). 

\begin{figure}[htbp]
\begin{center}
\includegraphics[width=60mm]{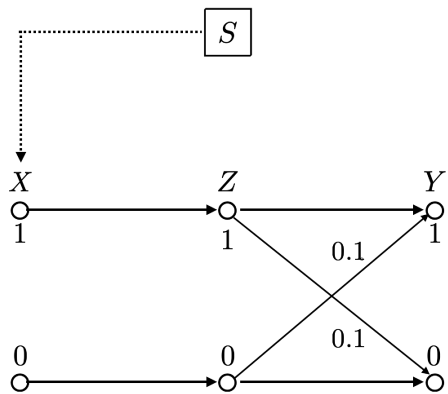}
\end{center}
\caption{ Reversely degraded binary  WTC with  causal CSI $S$ available only at Alice.}
\label{fig7}
\end{figure}

%
%
1) First, consider the reversely degraded binary WTC as in Fig.\ref{fig7} with $W(y,z|x,s) =W(y,z|x)$ with binary entropy 
$H(S) =1-h(0.2)<1-h(0.1)$.
It is easy to check that $I(SV;Y)-I(SV;Z)$ in (\ref{eq:silk-20q}) can be rewritten as 
\beqn\label{eq:dame-moto-10}
\lefteqn{I(SV;Y)-I(SV;Z)}\non\\
&=& (I(V;Y)-I(V;Z)) + H(S|UZ)-H(S|UY).\non\\
&&
\eeqn
Suppose here that $I(V;Y) =0$,  then
\beq\label{eq:tendo2}
I(S V;Y)=I(S;Y|V)=H(S|V)-H(S|VY),
\eeq
from which, together with   the constraint $I(S V;Y)\ge H(S)$  in (\ref{eq:silk-20q}), it follows that $H(S|V)-H(S|VY)\ge H(S),$
i.e., $-H(S|VY)\ge I(S;V)=0$ (owing to the independence of $S$ and $V$) and hence $H(S|VY)=0$.
On the other hand, in view of the Markov chain property $SV\to Z\to Y$ (due to the reverse degradedness) as well as  the 
independence of $S$ and $V$, it must hold that $H(S|VY) >0$
in that we are here considering the causal WC with CSI $S$ available only at Alice. This is a contradiction. 
Thus,  it should hold that  $I(V;Y) >0$ 
 for all $V$ satisfying the constraint $I(S V;Y)\ge H(S)$ and hence 
 $I(V;Y) \ge c_0$   for some $c_0>0$ and
  for all $V$ satisfying the constraint. 
  Furthermore, we can show also that  $I(V;Z)-I(V;Y)=I(V;Z|Y) >0$. To see this, assume $I(V;Z|Y) =0$ to lead to a contradiction.
  Then, it is easy to check that $I(V;Z|Y) =0$ together with  the reverse degradedness implies that $V$ and $ZY$ are independent and hence
  particularly $I(V;Y)=0$,
  which is a contradiction.
  Thus, 
  $ I(V;Y)-I(V;Z) \le -d_0$ for some $d_0>0$ and for all $V$ satisfying the constraint,  which,
  together with (\ref{eq:silk-20q}), (\ref{eq:riri7q}) and (\ref{eq:dame-moto-10}),
   implies that 
  \beq\label{eq:lkj1}
  K_1<K_2, 
  \eeq
  where we have taken account that the constraint
  in (\ref{eq:silk-20q}) is tighter than that in (\ref{eq:riri7q}).

\begin{figure}[htbp]
\begin{center}
\includegraphics[width=60mm]{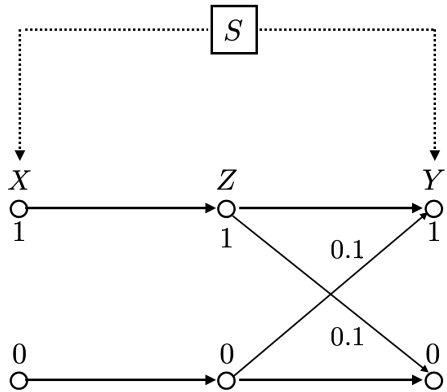}
\end{center}
\caption{ Reversely degraded binary  WTC with  causal CSI $S$ available at Alice and Bob.}
\label{fig5}
\end{figure}

2) Now consider the reversely degraded binary WTC as in Fig.\ref{fig5} with $W(y,z|x,s) =W(y,z|x)$. Then, setting 
$U=X$ ($\Pr\{X=1\}=\Pr\{X=0\}=1/2$)  independently of $S$ with  binary entropy $H(S)=1-h(0.1)$ in (\ref{hasa-2q1}), we have 
\beq
M_2^{\prime} \ge 1-h(0.1) > M_1^{\prime},
\label{eqq:p1}
\eeq
where  the first inequality  in (\ref{eqq:p1}) follows directly by observing that
in this case $H(S|UZ)=I(U;Y|S)=1-h(0.1)$ and  
 the second inequality  in (\ref{eqq:p1})
can be verified as follows. 
Suppose otherwise, i.e.,
\beq\label{eqq:as1}
1-h(0.1) \le M_1^{\prime}, 
\eeq
which then, together with  (\ref{eq:silk-10q1}),  means that there exists some $VX$ such that 
\beq\label{eq:kyou1}
I(V;Y|S) \ge 1-h(0.1),
\eeq
\beq\label{eq:kyou2}
I(V;Y|S)-I(V;Z|S) +H(S) \ge 1-h(0.1).
\eeq
On the other hand, (\ref{eq:kyou1}) implies that it must hold that $V=X$ 
($\Pr\{X=1\}=\Pr\{X=0\}=1/2$) independently of $S$  with $H(S)=1-h(0.1)$. Thus,
in view of (\ref{eq:kyou2}) with this $V=X$, we must have
\beq\label{eqq:as12}
I(X;Y|S)-I(X;Z|S) +H(S) \ge 1-h(0.1),
\eeq
which, together with $I(X;Y|S)=1-h(0.1), I(X;Z|S)=1$,  means that
\beq\label{eqq:as13}
-h(0.1) +H(S)\ge 1-h(0.1),
\eeq
that is, 
\beq\label{eqq:as14}
H(S)=1-h(0.1) \ge1,
\eeq
which is a contradiction, thus establishing the second inequality in (\ref{eqq:p1}).
%

%

%
%

%


%

\begin{figure}[htbp]
\begin{center}
\includegraphics[width=60mm]{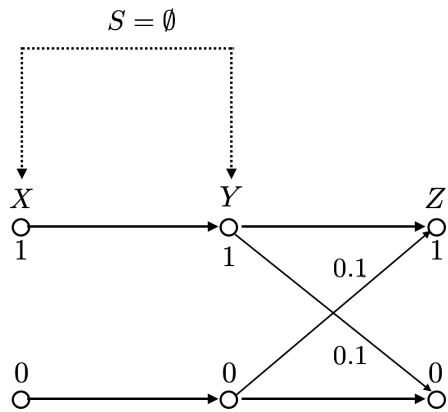}
\end{center}
\caption{Degraded binary WTC without  CSI  at Alice and Bob.
          }
\label{fig6}
\end{figure}

3) We next consider the degraded binary WTC as in Fig.\ref{fig6} with $W(y,z|x,s) =W(y,z|x)$ and $S=\emptyset$. 
Let $\Pr\{X=1\}=\Pr\{X=0\}=1/2$.
Then, since $S=\emptyset$,
(\ref{eq:silk-10q1}) and (\ref{hasa-2q1}) are evaluated  as 
\beqn
M_1^{\prime} &=& \max_{p_V}\min (I(V;Y)-I(V;Z), I(V;Y)),\non\\
&=&  \max_{p_V} (I(V;Y) -I(V;Z))\non\\
&\ge& I(X;Y)-I(X;Z) = h(0.1),
\label{eq:silk-10q1er}\\
M_2^{\prime} &=&0.
\label{eqq:p1rer}
\eeqn
Hence,
\beq
M_1^{\prime} > M_2^{\prime}.\label{eqq:sati-wr}
\eeq
Similarly, again by 
letting $\Pr\{X=1\}=\Pr\{X=0\}=1/2$ and taking account of  $S=\emptyset $, 
we see that (\ref{eq:silk-20q1}) reduces to
\beqn
K_1^{\prime} &\ge &  \max_{p_V}
 (I(V;Y)-I(V;Z))\non\\
 &\ge &  I(X;Y)-I(X;Z) \non\\
 &=& h(0.1) \non\\
 &>& 0 = K_2^{\prime},\non
\eeqn
that is,
\beq
K_1^{\prime}  > K_2^{\prime}. \label{eqq:mnb}
\eeq
Finally, summarizing up (\ref{eq:lkj1}), (\ref{eqq:p1}), (\ref{eqq:sati-wr}) and (\ref{eqq:mnb}),  the claim of Remark \ref{chui:sune3}
has been proved.
\IEEEQED
%

\section*{Acknowledgments}

The authors are grateful to Hiroyuki Endo for useful discussions. Special thanks go to the reviewers
for their stimulating comments which have occasioned to largely
improve  the quality of the earlier version.
This work was funded:
by JSPS KAKENHI Grant Number 17H01281, and partly supported by ``Research and 
Development of Quantum Cryptographic Technologies for Satellite Communications (JPJ007462)” 
of Ministry of Internal Affairs and Communication (MIC), Japan.

%

%

%
%

%


\bigskip

{\bf Te Sun Han} (M'80-SM'88-F'90-LF'11) received the B.Eng., M.Eng., and
D.Eng. degrees in mathematical engineering from the University of Tokyo,
Japan, in 1964, 1966, and 1971, respectively. Since 1993, he has been a
Professor with the University of Electro-Communications, where he has been a
Professor Emeritus since 2007. He has published papers on information theory,
most of which appeared in the IEEE TRANSACTIONS ON INFORMATION
THEORY. Also, he has published two books: one of them is Information-
Spectrum Methods in Information Theory (Springer Verlag, 2003). Especially,
this book was written to try to demonstrate part of the general logics latent in
information theory. His research interests include basic problems in Shannon
theory, multi-user source/channel coding systems, multiterminal hypothesis testing
and parameter estimation under data compression, large-deviation
approach to information-theoretic problems, and especially, information spectrum
theory. He is a member of the Board of Governors for the IEEE
Information Theory Society. He has been an IEICE Fellow and an Honorary
Member since 2011. He was a recipient of the 2010 Shannon Award. From
1994 to 1997, he was the Associate Editor for Shannon Theory of the IEEE
TRANSACTIONS ON INFORMATION THEORY.
\bigskip

{\bf Masahide Sasaki} received the B.S., M.S., and Ph.D. degrees in physics
from Tohoku University, Sendai Japan, in 1986, 1988 and 1992, respectively.
During 1992–1996, he worked on the development of semiconductor devices
in Nippon-Kokan Company (currently JFE Holdings). In 1996, he joined the
Communications Research Laboratory, Ministry of Posts and Telecommunications
(since 2004, National Institute of Information and Communications
Technology (NICT), Ministry of Internal Affairs and Communications).
He has been working on quantum optics, quantum communication and
quantum cryptography. He is presently Distinguished Researcher of Advanced
ICT Research Institute, and NICT Fellow. Dr. Sasaki is a member of
Japanese Society of Physics, and the Institute of Electronics, Information and
Communication Engineers of Japan.


\begin{thebibliography}{999}
%


\bibitem{wyner-wire} A. D. Wyner, ``The wire-tap channel," 
 {\em Bell Syst. Tech. J.}, vol.54, pp.1355-1387, 1975


\bibitem{csis-kor-3rd} I. Csisz\'ar and J. K\"orner, ``Broadcast channels with confidential messages,"
 {\em IEEE Transactions  Information Theory}, vol.24, no.3, pp.339-348, 1978
 

 
\bibitem{shannon-secrecy} C. E. Shannon, ``Communication theory of secrecy systems,"
 {\em  Bell Syst. Tech. J.},  vol. 28, pp.656-715, 1949
 
 \bibitem{shannon-cent} C. E. Shannon, ``Channels with side information at the transmitter,"
{\em  IBM J. Tech. Develop.},  vol. 2, no. 4, pp.289-293, 1958

 
\bibitem{mitrpant}  C. Mitrpant, A. J. H. Vink and Y. Luo, ``An achievable region for the Gaussian wiretap channel with side information," {\em IEEE Transactions on Information Theory,} vol. 52, no. 5, pp. 2181-
2190, 2006

\bibitem{chen-vinck} Y. Chen and A. J. H. Vinck, ``Wiretap channel with side information,"
{\em IEEE International Symposium on Information Theory}, Seattle, USA, July 2006;
{\em IEEE Transactions on Information Theory,} vol. 54, no. 1, pp. 395-402, 2008


 \bibitem{liu-chen} W. Liu and B. Chen, ``Wiretap channel with two-sided channel state information."
 {\em 41st Asilomar Conference on Signals, Systems and Computation,} November, 2007
 
 
 \bibitem{dai-vinck} B. Dai, Z. Zhuang and A. J. H. Vinck, ``Some new results on the wiretap
channel with causal side information,"{\em Proc. IEEE. ICCT,} Chengdu,
China, pp. 609-614, Nov. 2012


\bibitem{boche} H. Boche and R. F. Schaefer, ``Wiretap channels with side information-
strong secrecy capacity and optimal transceiver design," {\em IEEE Transactions on information
Forensics and Security,} vol. 8, no. 8, pp. 1397-1408, 2013

%
 
 %


  \bibitem{khisti-wornell-seoul} A. Khisti, S. Diggavi and G.Wornell,
``Secret Key Agreement Using Asymmetry in
Channel State Knowledge," {\em Pro. IEEE International Symposium on Information Theory,}
pp. 2286-2290, Seoul, 2009

 \bibitem{khisti-wornell} A. Khisti, S. Diggavi and G.Wornell, 
 ``Secret-key agreement with channel state information at the transmitter,"
{\em  IEEE Transactions on Information Forensics and Security}, 
no.3, vol.6, pp.672-681, 2011
 

 
 \bibitem{chia-elgamal} Y.  K. Chia and A. El Gamal, ``Wiretap channel with causal state information,"
 {\em  IEEE Transactions on Information Theory}, vol.IT-50, no.5, pp.2838-2849, 2012
 
  
  



 \bibitem{fujita} H. Fujita,  ``On the secrecy capacity of wiretap channels with
side information at the transmitter,"  {\em  IEEE Transactions on Information Forensics and 
Security}, vol.11, no.11, pp.2441-2452, 2016


 \bibitem{han-sasaki-c} T. S. Han and M. Sasaki,
 ``Wiretap channels with causal state information: strong secrecy,"
{\em IEEE Transactions on Information Theory}, vol.65, no.10, pp. 6750-6765, 2019 

 


 
  \bibitem{han-et} T. S. Han, H. Endo and M. Sasaki, ``Wiretap channels with one-time state information: 
  strong secrecy,"  {\em IEEE Transactions on Information Forensics and Security},
        vol.13, no.1, pp.224-236, 2018

 
 
  \bibitem{abc}  R. Ahlswede and I. Csisz\'ar,  ``Common randomness in information theory and cryptography-Part II: 
  CR capacity,"  {\em IEEE Transactions on Information Theory},
      vol. 44, no.1, pp.225-240, 1998


 
 
 
 \bibitem{prabhakaran} V. M. Prabhakaran, K. Eswaran and K. Ramchandran, ``Secrecy via Sources and Channels," {\em IEEE Transactions on  Information Theory,}  vol. 58, no.11, pp. 6747-6765,  2012



 
 
 
 \bibitem{dai-luo} B. Dai and Y. Luo, ``Some new results on the wiretap channel with side
information, {\em Entropy,} vol. 14, no. 9, pp. 1671-1702, 2012.


\bibitem{gamal-kim} A. El Gamal and Y.H. Kim, {\it Network Information Theory,}
Cambridge University Press , New York, 2011%


\bibitem{eva} E. Song, P. Cuff and V. Poor,
"``The likelihood encoder for lossy compression,"
{\em IEEE Transactions on  Information Theory,}  vol. 62, no.4, pp. 1836-1849,  2016

\bibitem{gel-pin} S. I. Gelfand and M. S. Pinsker, ``Coding for channel with random parameters,"
{\em Problems of Control and Information Theory},
vol.9, no.1, pp. 19-31, 1980


\bibitem{cuff-s}  P. Cuff,
``Strong soft-covering and applications,"
ArXiv:1508.01602v1, 2015

 \bibitem{goldfeld} Z. Goldfeld, P. Cuff, and H. H. Permuter,  
``Wiretap channel with random states non-causally available
at the encoder," {\em IEEE Transactions on  Information Theory,}  vol. 66, no.3, pp. 1497-1519, 2020
 
 \bibitem{bunin} A. Bunin, Z. Goldfeld, H. Permuter, S. Shamai, P. Cuff and P. Piantanida,
 ``Semantically-secured message-key trade-off over wiretap channels with random parameters,"
{\em Proc. of the 2nd Workshop on Communication Security}, pp. 33-48, 2018
  
  \bibitem{bunin-2019} A. Bunin, Z. Goldfeld, H. Permuter, S. Shamai, P. Cuff and P. Piantanida,"
 ``Key and  message semantic-security over state-dependent channels,"
{\em IEEE Transactions on Information Forensics and Security}, vol.15, pp. 1541-1556, 2020 

 \bibitem{csis-kor-2nd} I. Csisz\'ar and J. K\"orner, {\em Information Theory: 
Coding Theorems for Discrete Memoryless Systems}, 2nd ed., Cambridge University Press, 2011
John Wiley $\&$ Sons, NJ, 1968

\bibitem{gall} R. G. Gallager, {\it Information Theory and Reliable Communication,}
John Wiley $\&$ Sons, NJ, 1968




%
\bibitem{ver-han} T. S. Han and S. Verd\'u, ``Approximation theory of output statistics,"
 {\em  IEEE Transactions on Information Theory}, vol.IT-399, no.3, pp. 752-772, 1993
 %

\bibitem{zibae} A. Zibaeenejad, ``
Key Generation Over Wiretap Models
With Non-Causal Side Information," 
 {\em  IEEE Transactions on Information Forensics and Security}, vol.10, no.7,  pp. 1456-1471, 2015
 
 \bibitem{jafar}
 Syed Ali Jafar, ``Capacity with causal and non-causal side information: A unified view,"
  {\em IEEE Transactions on Information Theory,}
vol. 52, no. 12, pp. 5468-5474, Dec. 2006


 \end{thebibliography}
\end{document}